\documentclass[a4paper,fleqn,usenatbib]{mnras}
\usepackage{newtxtext,newtxmath}
\usepackage[T1]{fontenc}
\usepackage{ae,aecompl}
\usepackage{graphicx}	
\usepackage{amsmath}	
\usepackage{newtxtext,newtxmath}
\usepackage{multicol}        
\usepackage{bm}	
\usepackage{pdflscape}
\usepackage[usenames]{color}
\usepackage{soul}
\usepackage{wrapfig}
\usepackage[utf8]{inputenc}
\usepackage[section]{placeins}
\usepackage{float}
\usepackage{pgfplots}
\usepackage{natbib}
\usepackage[nottoc]{tocbibind}
\usepackage{color,soul}
\usepackage{enumerate}

\author[M. Gogilashvili et al.]{
Mariam Gogilashvili,$^{1}$ \thanks{Email: mg18u@my.fsu.edu}
Jeremiah W. Murphy,$^{1}$ \thanks{Email: jwmurphy@fsu.edu}
Quintin Mabanta$^{1}$
\\
$^{1}$Department of Physics, Florida State University, 77 Chieftan Way, Tallahassee, FL 32306, USA}

\title{Explosion Energies for Core-collapse Supernovae I:
  Analytic, Spherically Symmetric Solutions}

\begin{document}

\label{firstpage}
\pagerange{\pageref{firstpage}--\pageref{lastpage}}
\maketitle
\begin{abstract}
Recent multi-dimensional simulations of core-collapse supernovae are producing successful
  explosions and explosion-energy predictions.  In general, the
  explosion-energy evolution is monotonic and relatively
  smooth, suggesting a possible analytic solution.  We
derive analytic solutions for the expansion
of the gain region under the following assumptions: 
spherical symmetry, one-zone shell, and powered by neutrinos
and $\alpha$ particle recombination.  We consider two hypotheses: I) explosion energy
  is powered by neutrinos and $\alpha$ recombination,
   II) explosion energy is powered by neutrinos alone.  
 Under these assumptions, we derive the fundamental
  dimensionless parameters and analytic scalings. 
  For the neutrino-only hypothesis (II), the
  asymptotic explosion energy scales as 
$E_{\infty} \approx 1.5 M_g\varv_0^2 \eta^{2/3}$, where $M_g$ is the gain mass, $\varv_0$ is the free-fall velocity at
  the shock, and $\eta$ is a ratio of the heating and dynamical time
  scales.  Including both neutrinos and recombination (hypothesis I), the
  asymptotic explosion energy is $E_{\infty} \approx M_g
  \varv_0^2 (1.5\eta^{2/3} + \beta f(\rho_0))$, where $\beta$ is
	the dimensionless recombination parameter.
We use
 Bayesian inference to fit these analytic models to
   simulations.  Both hypotheses fit the simulations of
  the lowest progenitor masses that tend to explode spherically.  The fits do not prefer hypothesis I or II;
  however, prior investigations suggest that $\alpha$ recombination is
  important.  As expected, neither hypothesis fits the
  higher-mass simulations that exhibit aspherical explosions.  In
  summary, this explosion-energy theory is consistent with the spherical explosions of
  low progenitor masses; the inconsistency with higher
  progenitor-mass simulations suggests that a theory
  for them must include aspherical dynamics.
\end{abstract}

\begingroup
\let\clearpage\relax
\endgroup
\newpage

\section{Introduction}

The final fate  for most massive stars
is a core-collapse supernova
(CCSN). As one of the most energetic transients in the
  Universe, CCSNe are  important for a wide range of astrophysical origins
and processes.  If the explosion succeeds, then they leave
  behind a neutron star; if it fails, then the star collapses to a black
  hole.  CCSNe are major contributors to nucleosynthesis \citep{woosley2002}.  The kinetic energy released is a major contributor
  to turbulence in the ISM, and is likely a major component of stellar
evolution feedback \citep{low2004}. Their importance have inspired theorists and
  simulators for decades, and recent advancements in multi-dimensional
  simulations are leading to successful explosions and
  explosion energy evolution curves.  In this manuscript, we develop an analytic
theory to explain the explosion energy evolution of these
multi-dimensional simulations.

The fundamental challenges of core-collapse supernova theory are to
explain how collapse reverses into explosion, predict which stars will
explode, and to predict explosion energies, neutron star masses, etc.
Just before explosion, the Fe core collapses, bounces at nuclear
densities, and launches a shock wave that then stalls into an
accretion shock.  The shock stalls due to a combination of Fe
disassociation at the shock, neutrino losses, and the ram pressure of
the collapsing star.  If the star is to explode, a significant fraction of
the binding energy of the protoneutron star must be transported out to
the region just behind the shock.  \citet{colgate66} and \citet{wilson85} showed that neutrinos are
instrumental in transporting thermal energy from the protoneutron star
to the postshock region; this neutrino transport of energy may then
relaunch the stalled shock into an explosion.  However, several decades
of one-dimensional neutrino transport and hydrodynamic simulations
show that this delayed neutrino mechanism works in spherically
symmetric simulations in only the lowest mass progenitors \citep{Liebendorfer2001a,Liebendorfer2001b,Liebendorfer2005,
Rampp2002,Buras2003,Buras2006,Thompson2003,Kitaura2006,Muller2017,
Radice2017}.

Multi-dimensional simulations succeed in exploding where
  one-dimensional simulations fail
  \citep{Benz1994,Herant1994,Burrows1995,Janka1995}. \citet{murphy2008} showed that
  the critical condition for explosion is 30\% lower in
  multi-dimensional simulations, and \citet{mabanta2018} showed that
  neutrino driven convection (and mostly turbulent dissipation) is responsible for this reduction.
  Multi-dimensional simulations which simulate both multi-dimensional
  neutrino transport and hydrodynamics are computationally expensive,
  of order 10 million cpu-hours for three-dimensional simulations.
  To date, there are only $\sim$20 simulations with successful
explosions and explosion energy predictions \citep{Lentz_2015,Bruenn_2006, Bruenn_2016,
  10.1093/mnras/stv1611, Muller2019, 10.1093/mnras/sty2585,
  2019ApJS..241....7S, Steiner_2013, Vartanyan2019}.

Simulations are important theoretical tools that provide guidance
  in the understanding of a complex, nonlinear process.  They also
  promise to provide accurate predictions.  On the other hand, their
  expense do not allow for large systematic studies,
  and their complexity make it difficult to understand the fundamental
  physics and conditions for explosion.  If possible, analytic
  investigations are both less computationally expensive and provide a
  deeper understanding of the theory.  In turn, this deeper
	understanding can help guide further investigations using limited
	resources available for simulations.

There are now $\sim 20$ multi-dimensional simulations that
successfully explode \citep{Lentz_2015,Bruenn_2006, Bruenn_2016,
  10.1093/mnras/stv1611, Muller2019, 10.1093/mnras/sty2585,
  2019ApJS..241....7S, Steiner_2013, Vartanyan2019}, and the explosion energy curves of
  these simulations exhibit two characteristics
  that encourage  analytic investigations.  For one, the explosion
  energies are systematically under-energetic compared to explosion
  energies inferred from observations \citep{murphy2019}.  An analytic investigation may
  provide a deeper understanding of why they differ.  Part of this
  deeper understanding would include identifying the important
  dimensionless parameters of the problem.  Knowing the important
  parameters will enable theoretical sensitivity studies.  For
  example, one could identify how much the neutrino luminosity,
  neutrino cross-section, etc. must
increase so that the simulations would match the observations.
The second characteristic that encourages analytic investigations
is that explosion energy curves show a smooth monotonic increase with time.  This
suggests that one may develop an analytic or semi-analytic theory
describing the explosion energy of CCSNe. Our goal is to
derive analytic scalings for the explosion energy evolution, and use
these scalings to understand the sensitivity of the explosion energy
to the important parameters of the problem.

There have been attempts to derive an analytic expression for
explosion energy.  \cite{Janka2001} considered the
  delayed-neutrino mechanism and developed a toy model based on the
approximate solution of the hydrodynamic equations.  In this model, there are
three layers: the cooling layer (bounded by neutrinosphere below
 and gain region above), the heating layer (between the gain radius and
the shock) , and the infall layer (outside the shock).  The two
  lower layers are assumed
to be in hydrostatic equilibrium. They solved a boundary value
  problem at the shock and derived explosion conditions.
They found that if the star explodes,
  then the energy of the gain region can be at most $(2-3) \times
10^{51} erg$.  However, this study does not infer the
important scalings and parameters of the problem. 

In another analytic study,
\cite{Muller2016} also consider the delayed-neutrino mechanism
  and derive explosion energies based upon the structure of the
  progenitor.   In this model, the explosion phase is described by simple ODEs that take
into account continuous accretion after shock revival. In the
pre-explosion phase, they considered  a
layered model similar to \cite{Janka2001}. To simplify the
  calculations, they use a  one-zone approximation in the
explosion phase. This model assumes that the
explosion energy is  mostly due to
recombination energy. With these assumptions \citet{Muller2016}
derive an ODE  and numerically calculate
explosion energy  for a given stellar
structure and composition . This model
 produces explosions with energies up to 
$2\times 10^{51}$ erg. However, this study did  not solve the
ODE analytically.

In a third study,
 \cite{Papish2018} developed an order
  of magnitude analysis of the explosion energy evolution
. They conclude that explosions
with delayed-neutrino mechanism cannot give explosion energies 
$>0.5\times 10^{51} erg$. However, their explosion energy
analysis assumes a steep sensitivity on the gain radius: $E
  \propto r_{\rm gain}^{-3}$. Changing the radius of the gain region  by a factor of
$2$ changes the explosion energy by a factor of $10$. Given this
  steep sensitivity on one parameter, it is not
obvious that the delayed-neutrino mechanism does not work.

As a final remark, none of these previous studies compared their
derivations with multi-dimensional simulations.  In this manuscript,
one of our primary goals is to test the predictions of the analytic
theory that we derive.  Specifically, we will use Bayesian fitting
techniques to infer values for the theoretical parameters.  We will
also assess the robustness of fit.  In this way, we will either rule
out the analytic theories or show that they are consistent with
multi-dimensional simulations.
 
Before further discussing our theory, assumptions, and procedure,
  we discuss the possible importance of Fe recombination in the
  explosion energy evolution.  Previous investigations suggest that neutrinos and/or Fe
  recombination are important in determining the explosion energy
  evolution.  One of the primary roles of neutrinos in the explosion
  is to transport thermal energy from the proto-neutron star outward
  in radius $\sim 150 km$ \citep{colgate66, wilson85}.  While the proto-neutron
star becomes more bound, there is an opportunity for the gain region and
overlying layers to become unbound.  Hence neutrinos certainly play
some role in the explosion dynamics.

Fe dissociation and recombination may also play an important role in the
  explosion energy evolution.  Before collapse, the core is
composed of bound (Fe) nuclei.  During collapse, thermal
photodissociation unbinds these nuclei into free neutrons, protons,
and alpha particles \citep{Fern2009}.  This occurs at around a temperature of $T
\sim 9$ MeV (the binding energy of Fe is 8.8 Mev per nucleon).  This
``loss'' of energy during collapse causes the collapsing material to
have a softer equation of state and become more bound.  Without
this loss of energy or softening of the equation of state, the initial
bounce shock might have exploded the star; however, Fe dissociation
and neutrino losses ensure that the shock stalls at least momentarily
before it revives into explosion. During explosion, the expanding material cools, and the free neutrons, protons, and
neutrons recombine to form, first alpha particles and then Fe-group
nuclei.  This recombination adds thermal energy back into the material.

It is unclear whether Fe recombination has a net effect on the
  explosion energy evolution.  Technically, every instance of Fe
  recombination (energy injection) started with Fe dissociation
  (energy loss).  In principle, the net effect of Fe dissociation
  and Fe recombination is zero.  However, the transport of energy by
  neutrinos may complicate this. \cite{Muller2016} noted that for low-energy
  two-dimensional explosions, the explosion energy was proportional
  to the total Fe recombination energy.  They suggest that explosion
  is initiated when neutrinos heat the gain region just enough to
  compensate for initial losses due to Fe dissociation.  They then suggest that the
  explosion energy is set by the mass of the gain region and the Fe
  recombination energy.  However, it is not clear whether this
  actually happens.  For example, \citet{murphy2017} show that the critical
  condition for explosion is satisfied when the gain region can no
  longer support hydrostatic equilibrium.  If this is connected to the
  original Fe dissociation, then it is not obvious.  Furthermore, it
  is not clear if the explosion energies of the most recent
  simulations follow the same correlation between Fe recombination
  energy and explosion energy.

Since it is not clear whether Fe recombination contributes to the
explosion energy, we consider two hypotheses when deriving analytic
explosion energy evolution.  The first hypothesis (I) is that both
neutrino power and Fe recombination contribute to the explosion
energy.  The second hypothesis (II) is that only neutrino power contributes to
net explosion energy because Fe dissociation and Fe recombination cancel.

Inspired by the need for a deeper understanding and the smooth
monotonic curves, we use a simple model to derive the explosion
energy curve. We develop a one-zone, one-dimensional
hydrodynamical model. In this
model, the gain region initially is in hydrostatic equilibrium. We
also assume that the delayed neutrino mechanism
 initiates the explosion. We assume that as the explosion
begins, the mass accretion rate through the gain region is
zero. Since this model is spherically symmetric, we expect our model
to successfully describe spherical explosions but fail at
  describing explosion dominated by aspherical motions.  Since lower
  mass stars mostly explode with spherical symmetry, and higher mass
  stars mostly explode aspherically, we expect our theory to be
  consistent with the explosion of lower mass stars and to be
  inconsistent with higher mass stars. Our final
  goal is to compare the analytic solutions to simulations. 

The structure of this manuscript is as follows.
 Section~\ref{sec2} presents our theoretical
  derivation.  First, we identify our assumptions and derive the basic
  dimensionless ODEs.  To derive the dimensionless equations, we
  propose important dimensionless parameters.  In addition, we
  derive analytic scalings, making predictions on how the explosion
  energy depends upon the dimensionless parameters. 
In section~\ref{sec3}, we compare the analytic theory with
existing multi-dimensional simulations. Using Bayesian analysis, we fit
our theory to simulations and determine which are actually good fits. Finally, in section~\ref{sec4}, we summarize our results,  discuss the
physical implications of our theoretical model, and suggest ways
  in which future simulations may disprove this analytic theory.

\section{A Simple Theory for CCSN-Explosion-Energy Evolution} \label{sec2}

To derive analytic scalings 
for the explosion energy evolution, we develop a
  one-zone model.  In this derivation, we start with the full
  hydrodynamics equations and are explicit about the assumptions in this
  one-zone model.  This method of approximation ensures that the model is self-consistent and
  qualitatively includes all relevant physics.  While this model will not be quantitatively
  accurate, it will illuminate important qualitative aspects of the
  theory.  For example, this approach leads to  a set of natural dimensionless
  parameters and scaling relations.

\begin{figure} 
\includegraphics[scale=0.5]{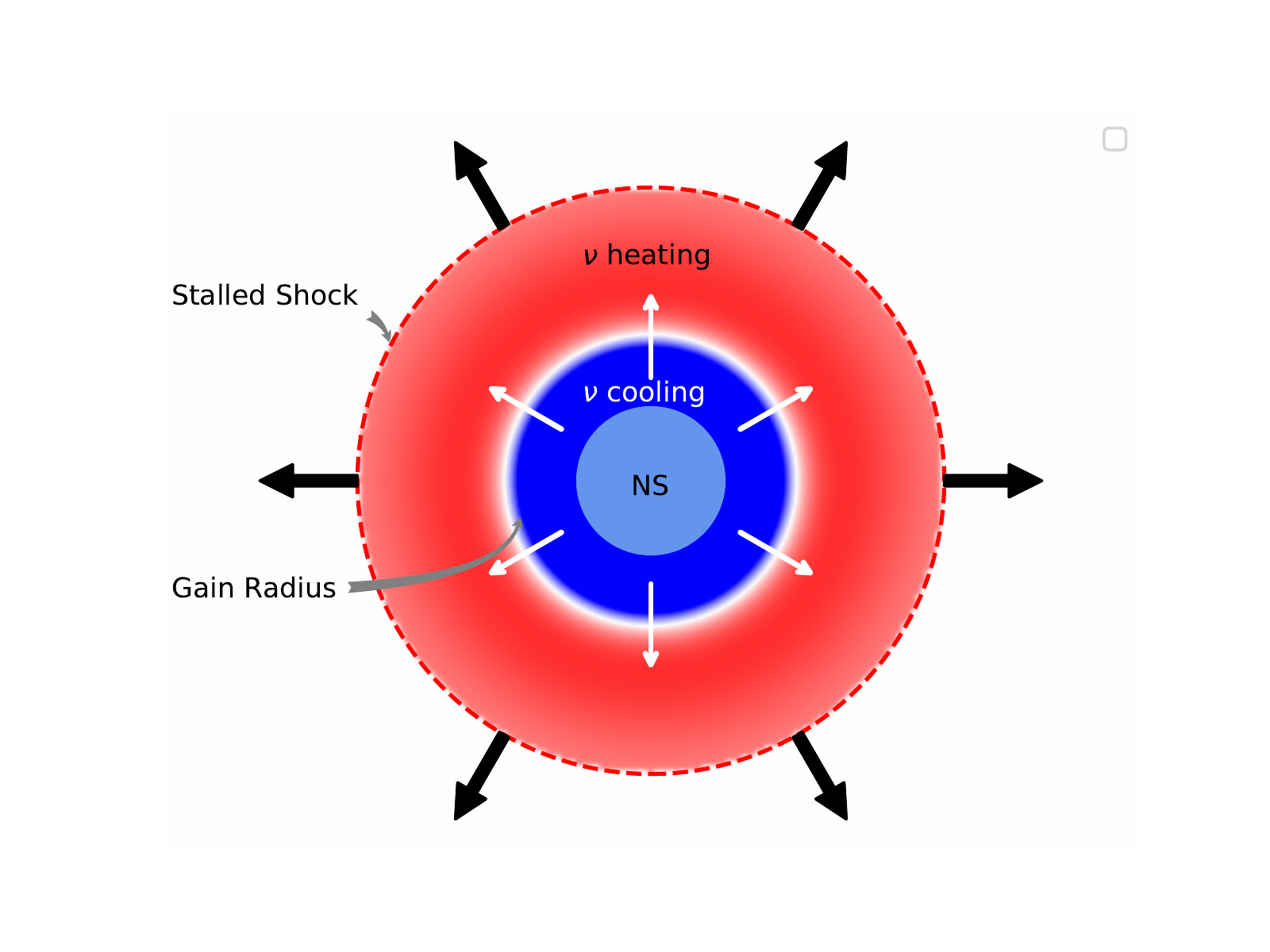}
\caption{
  Diagram for the spherical, one-zone explosion model.
   In this model, $t=0$ corresponds to the
	initiation of explosion: when postshock pressure begins to
	overwhelm gravity \citep{murphy2017}.  We model the energy evolution
	of the gain region in a one-zone approximation.   Neutrino
  heating accelerates material in the gain region outwards and it
  expands spherically. For initial conditions, the
	gain region is in hydrostatic balance with a neutrino power source
  below.} 
\label{ccsndiagram}
\end{figure}

Figure~\ref{ccsndiagram} shows a schematic of the one-zone model and
we use it to illustrate all assumptions of this model.  For one, we assume
spherical symmetry.  Second, we model the dynamic evolution of the
gain region.  As
CCSN simulations begin to explode, the gain region (where neutrino
heating dominates neutrino cooling) starts to become unbound and
expands \citep{wilson85}.  This gain region begins in near hydrostatic balance and
explosion begins when the structure can no longer maintain a
hydrostatic structure \citep{murphy2017}.  Hence a third major
  assumption is that the gain region
begins in hydrostatic balance.  Before explosion, during the stalled
phase, the mass accretion through the gain region is nonzero; as the region begins to explode,
mass accretion halts and reverses into explosion.  To simplify
  the parameters and equations, the fourth major assumption is  that the mass
accretion rate is zero through the gain region. 

The model also assumes that explosion is driven by neutrinos
  and/or $\alpha$ recombination. We consider two hypotheses:
\begin{enumerate}[I)]
\item  Explosion is powered by both neutrinos and $\alpha$
  recombination.  \cite{Muller2016} suggests that explosion initiates when
  neutrinos heat the gain region enough to compensate for the losses
  by Fe dissociation.   Thus,
  neutrinos only contribute to unbind the system. As a
  result, the system starts to explode from hydrostatic
  equilibrium and explosion energy is mostly set by Fe recombination.
  Later, we show that both neutrinos and
	$\alpha$ recombination can have a comparable impact on the
	explosion energy.

\item Explosion is powered by neutrinos alone.  When the core
  collapses at a temperature $T \sim 8.8 MeV$, Fe
  dissociates into nucleons and in doing so, the system
  becomes more bound by
  $\sim \epsilon_{\rm rec} M_g/m_p $. Later, in the explosion
  phase, nucleons again recombine to bound nuclei, which
   deposits approximately the same amount of
  energy $\sim \epsilon_{\rm rec} M_g/m_p$. If this is true, then Fe
	recombination has no effect on the explosion energy and only the
  neutrino power dictates the explosion energy.

\end{enumerate}

Our goal in this manuscript is not to
solve the full numerical solution but to derive
the important dimensionless parameters and analytic scaling relations that
govern the energy evolution for spherically symmetric explosions.
To
facilitate these analytic derivations, we assume a one-zone spherically
symmetric model for the dynamic evolution of the gain region. Within the
  context of this one-zone model, specific assumptions include:
 during explosion, the inner boundary is fixed
at the surface of the NS while the outer boundary of the gain
region expands with the shock; the fluid velocity is assumed to
be zero at the NS , and we assume
  that all expansion velocities scale by the same
  parameter; moreover, we assume that the ram pressure as well as
mass accretion rate is negligible in the dynamics of gain region.
For more details on
  these assumptions, see Appendix \ref{appendix}. 
 
\subsection{Fundamental Equations and Dimensionless Parameters} 

The hydrodynamics equations for conservation of mass, momentum, and
energy are:
\begin{equation} \label{continuity}
\frac{\partial \rho}{\partial t} + \boldsymbol{\nabla} \cdot (\rho
\boldsymbol{\varv})=0 \, ,
\end{equation}
\begin{equation} \label{Euler}
\rho \frac{d\boldsymbol{\varv}}{dt}=-\boldsymbol{\nabla} p -\rho
\boldsymbol{\nabla} \Phi \, ,
\end{equation}
and
\begin{equation} \label{InternalEnergy}
\rho \frac{d \epsilon}{dt}=-p \boldsymbol{\nabla} \cdot
\boldsymbol{\varv}+\rho q_{\nu} +\rho q_{\alpha} \, ,
\end{equation}
$\rho$ is the mass density, $\boldsymbol{\varv}$ is the velocity, $p$
is the pressure, $\epsilon$ is the specific internal energy, $\Phi$ is
the gravitaional potential, $q_{\nu}$ is the local specific neutrino heating
term, $q_{\alpha}$ is the
  change in internal energy due to $\alpha$ recombination.  To facilitate analytic solutions, we assume a polytropic
equation of state (EoS), $P = (\gamma - 1) \rho \epsilon$.
For gravity, we assume a simple Newtonian potentional, 
$\Phi=-\frac{G M_{NS}}{r}$, and for neutrino heating, a simple neutrino heating profile of $q_{\nu}=\frac{L_{\nu} \kappa}{4 \pi r^2}$.  The neutrino oppacity is
$\kappa=\frac{\sigma}{m_p}$. 

The difference in hypotheses I and II are the source terms
in eq.~(\ref{InternalEnergy}). For hypothesis I, we include
both the neutrino and $\alpha$ recombination terms.  For hypothesis II, we include only
  the term related to neutrino heating.  Within the gain region,
  neutrino cooling is much weaker than neutrino heating.  Therefore,
  we neglect neutrino cooling.  

To construct the dynamic equations for the spherical
shell, we average equations~(\ref{continuity})-(\ref{InternalEnergy})
over volume (see details in Appendix~\ref{appendix}). Using the Polytropic EoS, and the assumption that the
mass flow through the gain region is negligible at the initiation of explosion: 
\begin{equation} \label{dvdt}
\frac{d\varv}{dt}=-\frac{G M_{NS}}{r^2}+\frac{4 \pi r^2 p}{M_g},
\end{equation}
\begin{equation} \label{dpdt}
\frac{dp}{dt}=-3 \gamma \frac{p\varv}{r}+\frac{L_{\nu} \kappa (\gamma-1)}{\frac{16}{3} \pi^2 r^5} M_g + \frac{3M_g}{4\pi r^3}(\gamma-1) q_{\alpha} ,
\end{equation}
In these equations, $\varv$, $p$, and $r$ represent characteristic velocity,
pressure, and radius for the shell. Appendix \ref{appendix}
presents a thorough derivation of these equations; the
  appendix distinguishes the characteristic variables with a subscript ``c'';
  for convenience, we drop the subscript here. The last term in equation ~(\ref{dpdt}) vanishes for hypothesis II.

Again, for the initial conditions, we assume that the shell starts in
hydrostatic equilibrium. However, at the moment that the
evolution starts, there is no longer balance between
the pressure gradient and gravity. Therefore,
at t=0, we have the following initial conditions: $\varv=0$, $r=r_0$
and $p=p_0=\frac{G M_{NS} M_g}{4 \pi r_0^4}$. To derive this
This latter condition, assume $d\varv/dt = 0$ in
 eq.~(\ref{dvdt}) .

\begin{figure}
\includegraphics[scale=0.5]{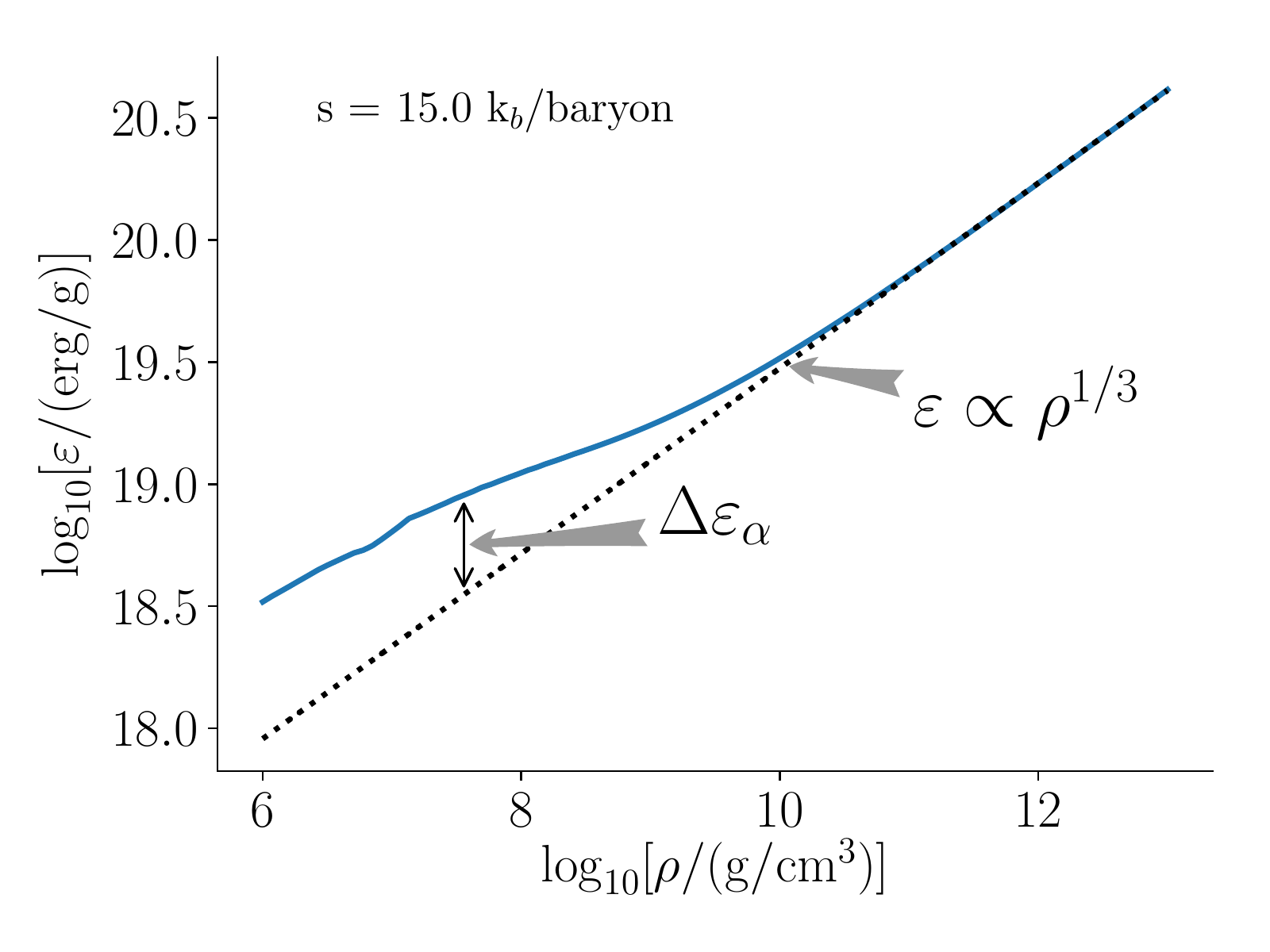}	
\caption{Specific internal energy ($\varepsilon$) vs. density
	$\rho$ for an adiabat with 
  $s=15 k_b/$baryon. We use the tabulated EOS
  of \citet{Steiner_2013}. This figure illustrates the
	estimate for the heating by $\alpha$ recombination.  For $\rho$
	around $10^{12}$ g cm$^{-3}$, the internal energy is dominated by photons and
	relativistic electrons and positrons.  Hence, $\epsilon\propto
	\rho^{1/3}$.  At lower densities, the free baryons start to
	combine to form $\alpha$s.  This raises the internal energy.  The
	difference between the actual internal energy and the
	extrapolation is a good estimate for the energy added by $\alpha$ recombination.  }
\label{EOS}
\end{figure}

To determine internal energy change due to $\alpha$
  recombination ($q_\alpha$), we note that the nuclear reaction rates
  are faster than the hydrodynamic rates.  Therefore, the $\alpha$
  fraction is in nuclear statistical equilibrium and is included in
  tabulated equations of state.  At sufficiently high densities and/or
  temperatures, the $\alpha$ fraction is zero.  Once material is
  heated to sufficiently high entropies, the material becomes unbound
  and expands away from the NS adiabatically.  Figure~\ref{EOS} shows
  an adiabat with $s = 15$ $k_{\rm{b}}$/baryon from the tabulated EOS of \citet{Steiner_2013}. Following this adiabat
  down in density, the baryons recombine to form $\alpha$s.  In the
  process, the recombination releases energy raising the internal
  energy.  We will use this to estimate the amount of energy liberated
  as a function of density.  

The first step is to estimate the extra amount of internal energy due
to the $\alpha$ recombination, $\Delta \varepsilon_{\alpha}$.  For the
region above the NS and below the shock, the EOS is dominated by
relativistic electrons and positrons, photons, free nucleons, and
$\alpha$s.  The partial pressure of the free nucleons and $\alpha$s
is described by a non-relativistic ideal gas.  The electrons,
positrons, and photons, which dominate the EOS are described by a
degenerate and relativistic EOS.  As a result, $\gamma \approx 4/3$.
Under this approximation, $\varepsilon \propto \rho^{1/3}$.  The
dotted line in Figure~\ref{EOS} is a fit at the highest densities, the
power-law index of this fit is 0.38.  This is as expected given that the
EOS is dominated by the relativistic species but also has some
contribution from the nonrelativistic species.  Extrapolating this fit
to lower densities and comparing to the actual internal energy gives
an estimate for how much internal energy is added due to $\alpha$
recombination, $\Delta \varepsilon_{\alpha}$.

The difference in the actual internal energy and the extrapolated
  roughly exhibits the following form
\begin{equation}
\Delta\epsilon_\alpha=a \tanh\left(\frac{11-log_{10}\rho}{b}\right) \,
,
\end{equation}
where $a=6.67 \times 10^{18} erg/g$ and $b=2.147$ are the
fitting parameters for the $s=20$ $k_{\rm{b}}$/baryon adiabat.  We
checked several other adiabats ($s= 5$, $10$, $15$, $20$ $k_{\rm{b}}/$baryon ) and found similar parameters. 
The heating rate due to $\alpha$ recombination, $q_{\alpha}$ is $\frac{3 a}{b \ln{(10)}}\frac{\varv}{r}\cosh^{-2}{\left(\frac{11-log_{10}\rho}{b}\right)}$ .

Defining the dimensionless parameters helps to illuminate the
important physics and scalings.  The dimensionless parameters of
  this model are as follows.   We scale the
radial coordinate by the initial position of the shell,
\begin{equation}
\tilde{r}=\frac{r}{r_0} ,
\end{equation}
the velocity by the dynamical velocity at $r_0$,
\begin{equation}
\tilde{\varv}=\frac{\varv}{\varv_0}= \varv {\left(\frac{G M_{NS}}{r_0}\right)}^{-1/2} ,
\end{equation}
the time coordinate by the dynamical time,
\begin{equation}
\tilde{t}=\frac{t}{t_0}=\frac{\varv_0}{r_0}t , 
\end{equation}
and the pressure by the initial hydrostatic pressure,
\begin{equation}
\tilde{p}=\frac{p}{p_0}.
\end{equation}
The resulting dimensionless coupled set of ordinary differential equations are:
\begin{equation} \label{drdttilde}
\frac{d\tilde{r}}{d\tilde{t}}=\tilde{\varv}
\end{equation}
\begin{equation} \label{dvdttilde}
\frac{d\tilde{\varv}}{d\tilde{t}}=-\frac{1}{\tilde{r}^2}+\tilde{p}\tilde{r}^2 ,
\end{equation}
\begin{equation} \label{dpdttilde}
\frac{d\tilde{p}}{d \tilde{t}}=-3\gamma \frac{\tilde{p}\tilde{\varv}}{\tilde{r}}+\frac{\eta}{\tilde{r}^5} + \frac{9 (\gamma-1)}{b \ln(10)} \beta \frac{\tilde{\varv}}{\tilde{r}^4} \cosh^{-2}\left( \frac{11-log_{10}(\rho_0/\tilde{r}^3)}{b} \right),
\end{equation}
where $\rho_0= \frac{3M_g/4 \pi r_0^3}{1 g/cm^3}$ is
  the initial density of the system.

Other than the dimensionless variables, these equations depend upon
three dimensionless parameters.  One is $\gamma$, the adiabatic index for
the EOS.  The gain region is dominated by free nucleons, photon, and
relativistic electrons and positrons.  Therefore, for the rest of this
manuscript, we consider $\gamma$ to be fixed at a value a little above
4/3.  The second dimensionless parameter is
\begin{equation} \label{eta}
\eta=\frac{t_0}{t_H}=\frac{3 L_{\nu}\kappa (\gamma-1)}{4\pi G M_{NS}
  \varv_0} \, ,
\end{equation}
which describes the ratio of dynamical time to heating time. The third dimensionless parameter is  
\begin{equation} \label{beta}
\beta=\frac{a}{G M_{NS}/r_0}
\end{equation}
which represents the maximum dimensionless internal energy change due to $\alpha$ recombination. For  hypothesis II, explosion is
dominated by neutrino heating, and $\beta=0$. 

In either
  case,  the total dimensionless energy of the shell is:
\begin{equation} \label{Etilde}
\tilde{E}(\tilde{t},\eta,\beta)=-\frac{1}{\tilde{r}}+\frac{\tilde{\varv}^2}{2}+\frac{\tilde{p}\tilde{r}^3}{3
  (\gamma-1)} \, .
\end{equation}
This represents the explosion energy in this simple one-zone model.
We also explicitly show that this dimensionless energy depends upon
the dimensionless time, $\tilde{t}$,  the heating parameter,
$\eta$, and the recombination parameter $\beta$.  In
  section ~\ref{2.3}, we show that $\eta$ determines the final dimensionless explosion
  energy for hypothesis II, and both $\eta$ and $\beta$ determine the
  final dimensionless explosion energy for hypothesis I.

The dimensionful explosion energy in either case is
\begin{equation} \label{E}
E=M_g \varv_0^2 \tilde{E}(\tilde{t},\eta,\beta) \, .
\end{equation}

\subsection{Perturbative Analysis}
Before we present solutions to
  equations~(\ref{drdttilde})-(\ref{dpdttilde}), we check the
  consistency of our model and equations with a perturbative
  analysis.  In the previous section, we made several major
  assumptions that could affect the consistency of the equations.  In
  particular, the one-zone assumption requires choices for
  divergences and variables that can affect the qualitative character
  of the evolution.  One important constraint is that when there is neither
external heating, $\eta = 0$, nor recombination, $\beta=0$,  then the zone should remain in
hydrostatic balance.  In other words, $dr/dt = 0$.  If the position is
perturbed, then it should oscillate about the hydrostatic position.
Therefore, we perturb the equations with $\eta = 0$ and $\beta =
  0$ to ensure that the
solutions are stable.

The decomposed variables are $\varv=\varv^{\prime}$, $r=r_0+r^{\prime}$,
  and $p=p_0+p^{\prime}$, and the resulting perturbed evolution equations are

\begin{equation}
\frac{d^2r^{\prime}}{dt^2}=\frac{4G M_{NS}}{r_0^3} r^{\prime}+\frac{4\pi r_0^2}{M_g}p^{\prime}
\, ,
\end{equation}
and
\begin{equation}
\frac{dp^{\prime}}{dt}=-3\gamma \frac{p_0}{r_0} \frac{dr^{\prime}}{dt} \, .
\end{equation}

One may combine these equations into one second-order equation for
$\varv^{\prime}$.  The solution of this second-order equation for
$\varv^{\prime}$ is proportional to $e^{i\omega t}$.  The resulting
expression for the frequency is

\begin{equation}
\omega^2=(3\gamma-4) \frac{G M_{NS}}{r_0^3} \, .
\end{equation}
As expected, solutions with $\gamma > 4/3$ oscillate about a
  hydrostatic equilibrium, and solutions with $\gamma < 4/3$
  correspond to either exponentially decaying or expanding solutions.
  Therefore, despite the large assumptions , the model at
  least produces the correct linear behavior.

\subsection{Solutions for Explosion Energy and Important Scalings}\label{2.3}

To solve the system of coupled differential equations
~(\ref{drdttilde})-(\ref{dpdttilde}), we use ODEINT from
the scipy.integrate package. (ODEINT solves differential equations using
LSODA from the FORTRAN library odepack.)

\begin{figure}
\includegraphics[scale=0.5]{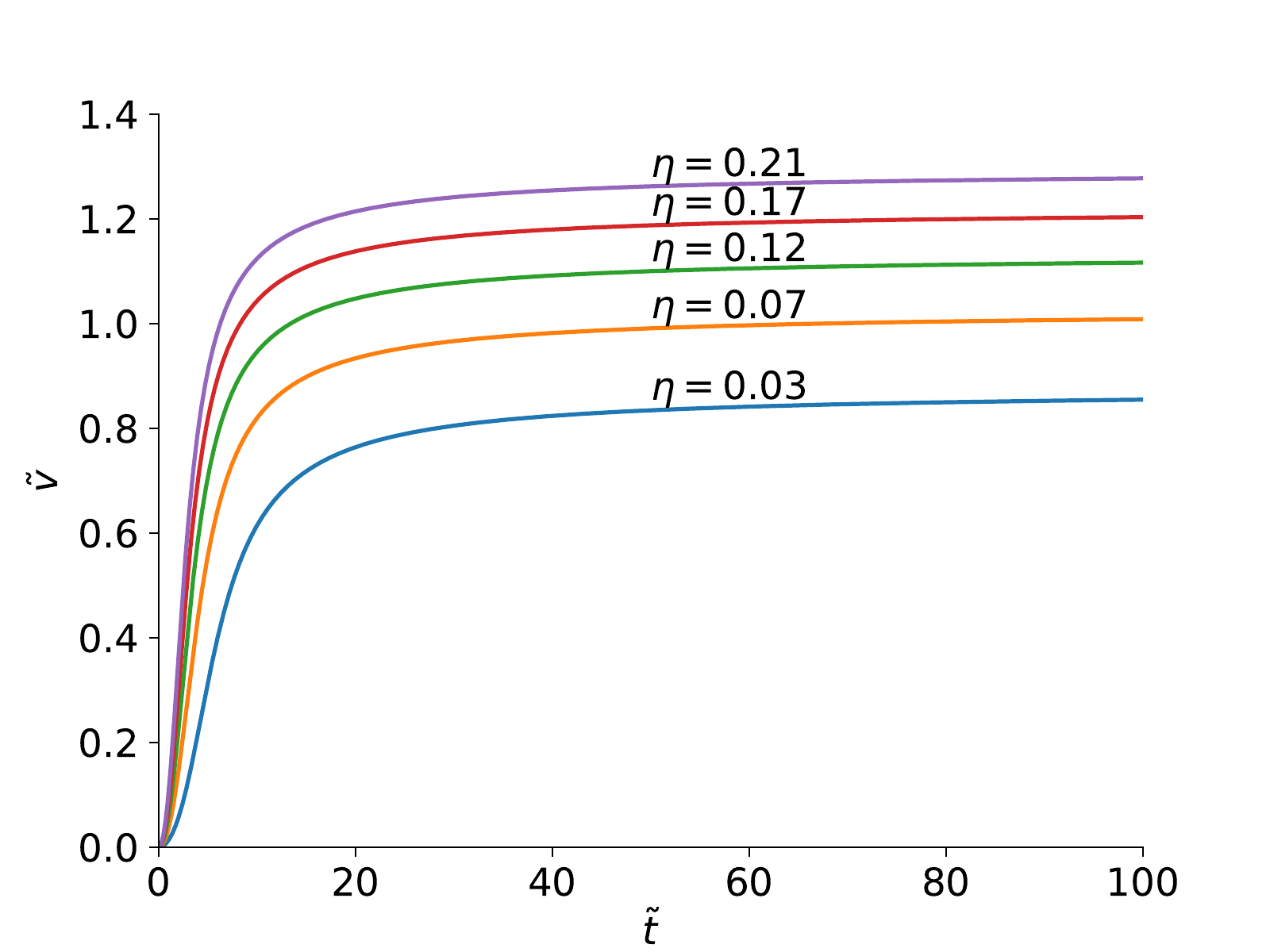}
\caption{Solutions for the dimensionless velocity
  $\tilde{\varv}$ as a function of dimensionless time ($\tilde{t}$) and
  heating parameter ($\eta$) for hypothesis I.  The other
	paramters are fixed at $r_0=150$km and $M_g=0.05$ M$_\odot$. The rise time of
  $\tilde{\varv}$ is mostly
  set by the dynamical timescale, while the
  value of velocity afer infinite time is mostly set by $\eta$
  and $\beta$.}
\label{Vteta}
\end{figure}

\begin{figure}
\includegraphics[scale=0.5]{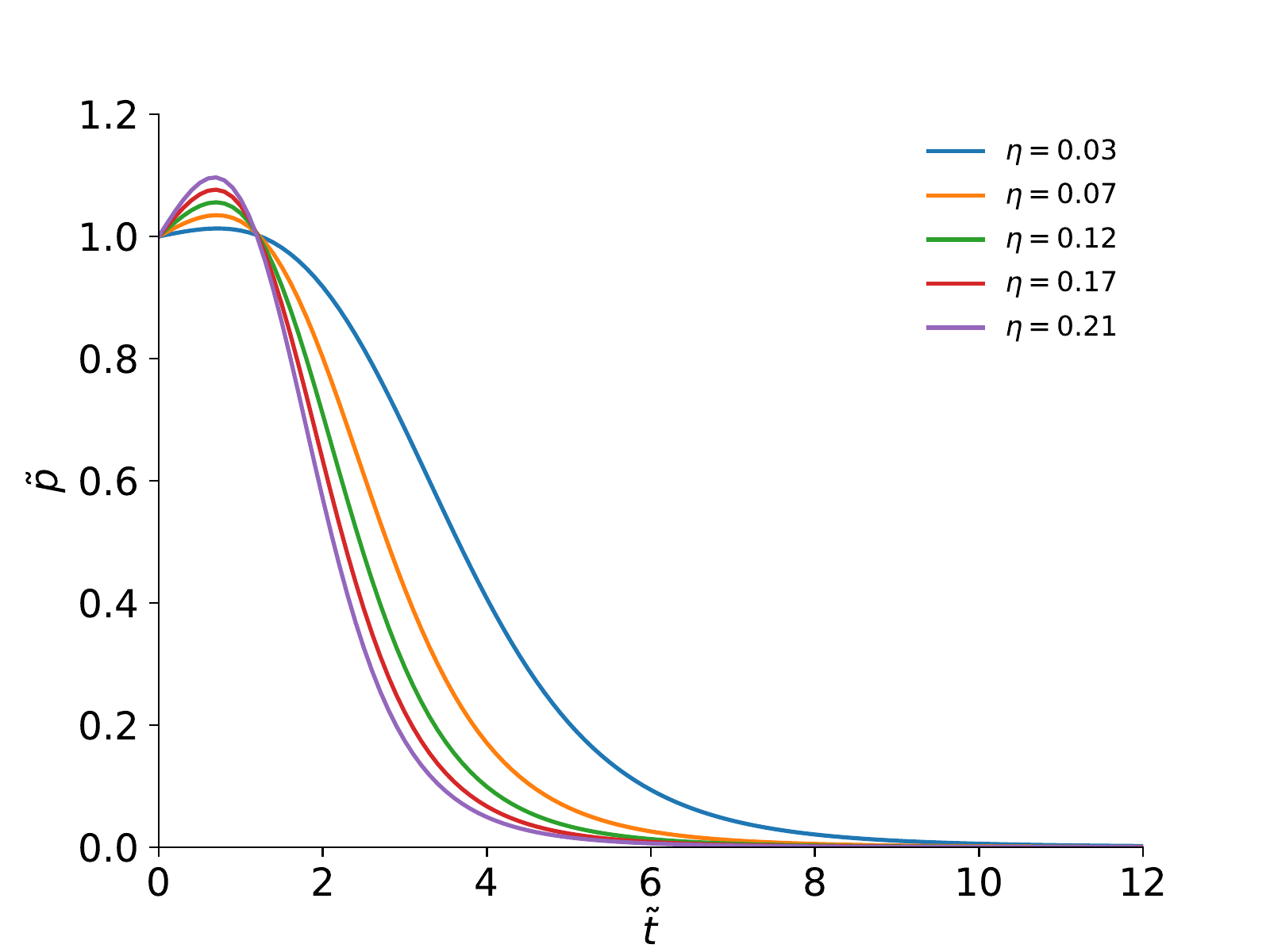}
\caption{Solutions for the dimensionless pressure $\tilde{p}$ as
  a function of dimensionless time and $\tilde{t}$ and heating
  parameter, $\eta$, for hypothesis I. The other parameters
	are fixed at $r_0=150$km and $M_g=0.05$ M$_\odot$. Until $\tilde{t} \approx
  2$, neutrino heating dominates and pressure
  increases. Afterwards, adiabatic cooling dominates and
  pressure decreases. The maximum value of pressure is set by
  the heating parameter $\eta$.}
\label{Pteta}
\end{figure}

Figures ~\ref{Vteta}~\&~\ref{Pteta} show
 the solutions for $\tilde{\varv}$ and $\tilde{p}$
as a function of time and the heating parameter,
$\eta$ for hypothesis I and for fixed values of $r_0=150 km$ and $M_g=0.05M_\odot$. The dynamical timescale, $t_0$, sets the rise time for
$\tilde{\varv}$. 
The heating parameter mostly sets the asymptotic value of 
$\tilde{\varv}$ . Neutrino heating dominates the earliest
  phase until
$\tilde{t}\approx 2$; neutrino heating falls off quite rapidly
  with radius, and as the gain region expands, it quickly leaves the
  source of neutrino heating.  Instead, as the gain layer expands,
  adiabatic cooling dominates. 

\begin{figure}
\includegraphics[scale=0.5]{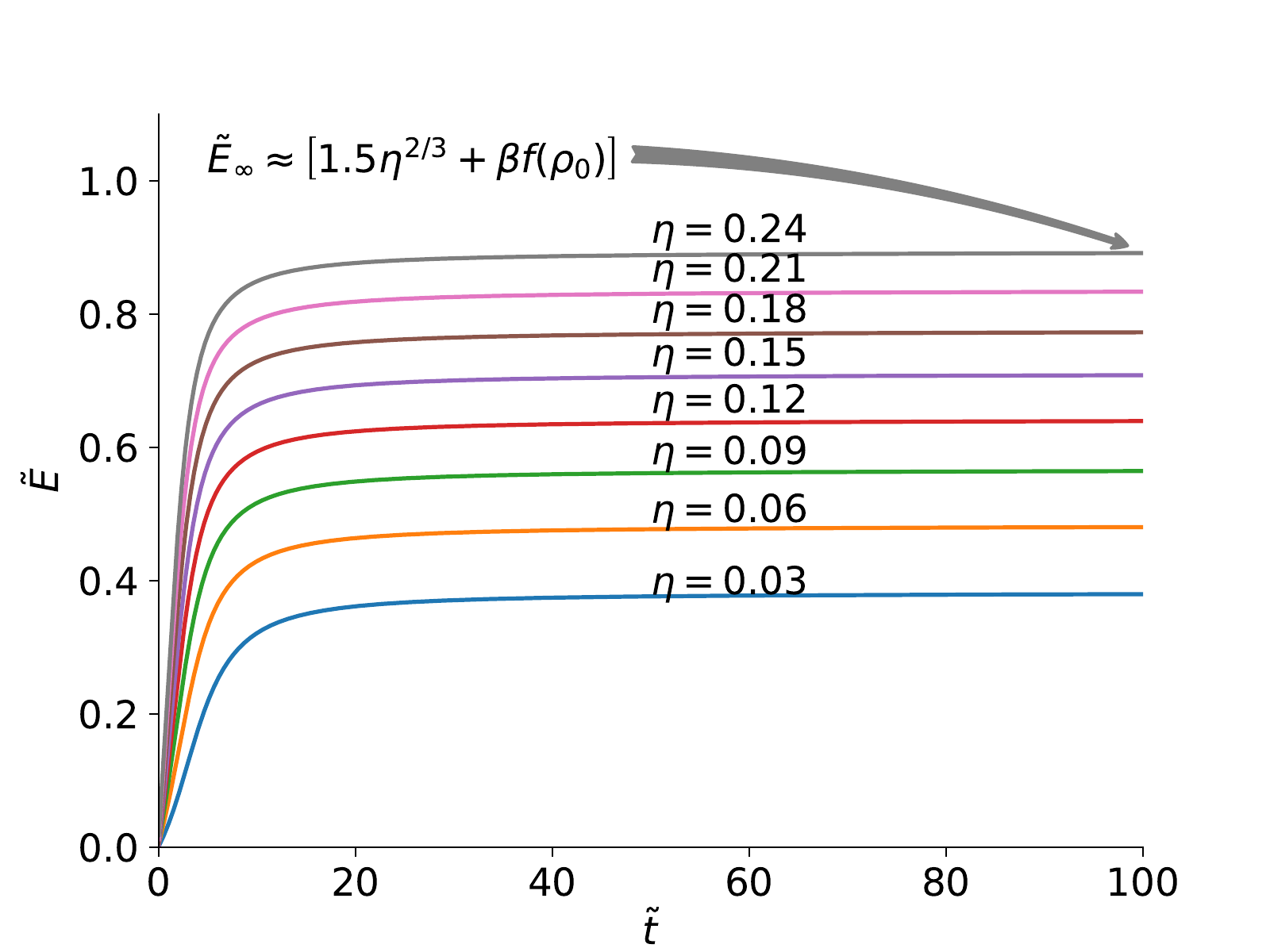}
\caption{ Dimensionless explosion energy
  $\tilde{E}(\tilde{t},\eta,\beta)$ as a function of dimensionless
  time and heating parameter $\eta$ for hypothesis I, for the fixed
  value of initial shock radius $r_0=150$km (i.e $\beta$).
   If $M_g=0.05$ M$_{\odot}$, $M_{NS}=1.4$ M$_{\odot}$, $r_0=150$ km and
  $\kappa=2\times10^{-17}$ cm$^2$ g$^{-1}$, then 
  $\eta = 0.12$ corresponds to a neutrino luminosity of $4.83 \times 10^{52}
	\frac{erg}{s}$. For most cases,
	the energy rises to 75\% of $\tilde{E}_{\infty}$ within
	$\tilde{t} = 3.5$.  This rise time is set by the dynamical time;
	the asymptotic explosion energy is set by $\eta$ and $\beta$; see eq.~(\ref{Einf}).}
\label{E_rec}
\end{figure}

 Figure ~\ref{E_rec}, 
represents the solutions for explosion energy in the case of
hypothesis I.  In this case, the
dimensionless explosion energy depends on three parameters:
dimensionless time, heating parameter $\eta$, and $\beta$.   As
  in hypothesis I, the explosion timescale is  mostly
determined by the dynamical timescale. For a fixed
  $\beta$, the explosion energy is set by $\eta$.   It may appear that including
  $\beta$ in the dynamics adds another parameter, increasing the
  number of independent parameters to 4.  However,   the
 dimensionful variables in $\beta$ are already in the other
  dimensionless parameters.  Therefore, there are still only three
  independent parameters.

\begin{figure}
\includegraphics[scale=0.5]{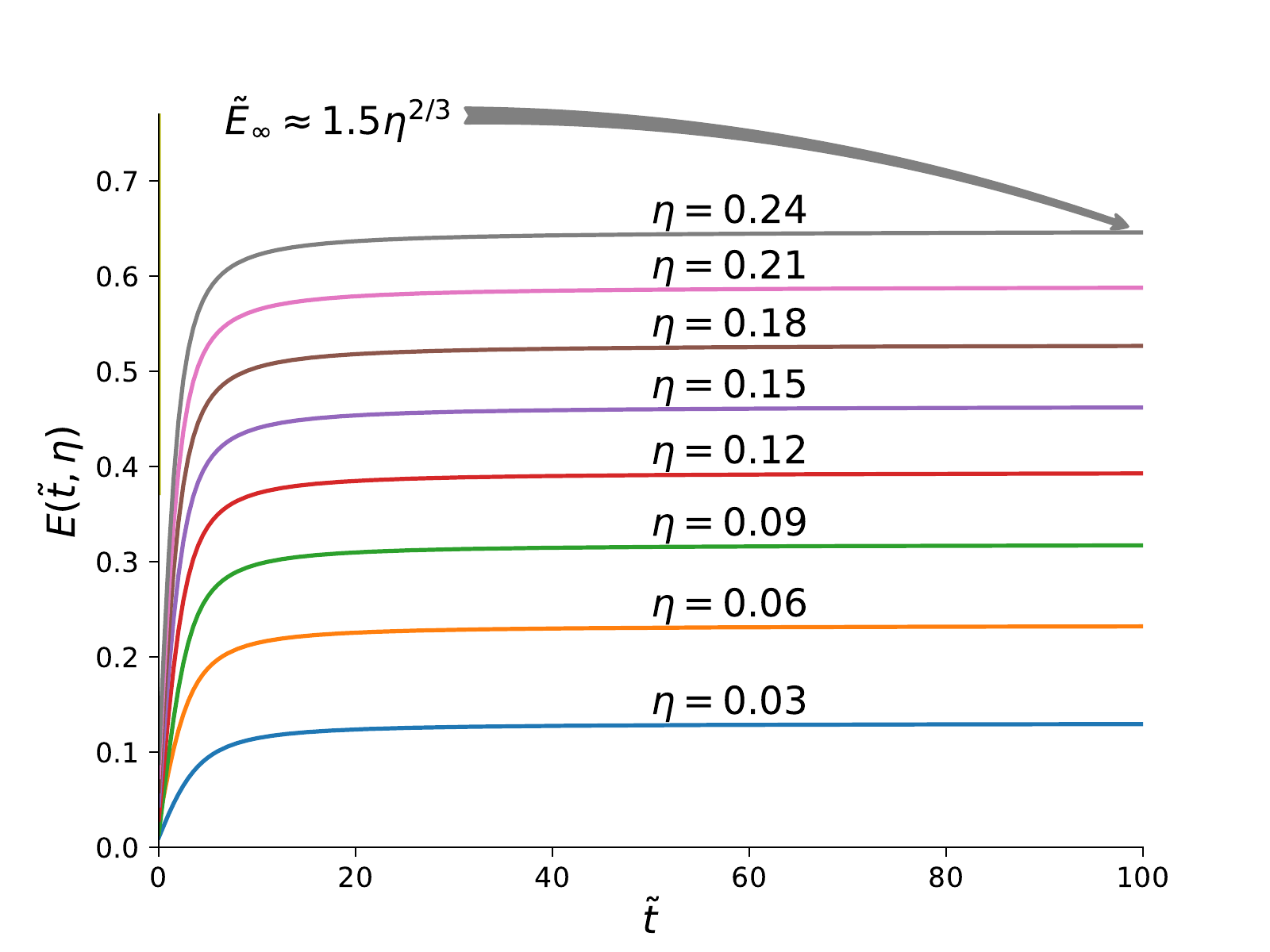}
\caption{Similar to Figure~\ref{E_rec} except for hypothesis
  II.  As in hypothesis I, the rise time is mostly determined by the
  dynamical time.  Since recombination offsets the dissociation during
  collapse for hypothesis II, the aymptotic energy is set by neutrinos ($\eta$)
  alone. }
\label{ETildevsTilde}
\end{figure}

Figure
~\ref{ETildevsTilde} shows the resulting 
solutions for dimensionless energy $\tilde{E}$ as a function of
$\tilde{t}$ and $\eta$ for hypothesis II. The
value of $\tilde{E}_{\infty}$ mostly depends on $\eta$, and
the rise time mostly depends on
the dynamical timescale, $t_0$.  

The explosion energy
  evolution depends upon three parameters.  Equation~(\ref{E}) expresses the
explosion energy in terms of mass in the gain region, $M_g$, the
dynamical velocity, $\varv_0$, and the dimensionless energy,
$\tilde{E}$.  In turn, $\tilde{E}$ depends upon the dynamical
timescale, $t_0$, and the heating parameter, $\eta$.  Upon first
glance, it may appear that the explosion energy evolution depends upon
four parameters.  However, $\varv_0$ and $t_0$ are not independent;
$t_0$ is proportional to $r_0/\varv_0$.  Furthermore, both $\varv_0$
and $\eta$ depend upon the neutron star mass, $M_{\text{NS}}$.

  In
summary, there are three independent parameters that determine the
explosion energy evolution.  In section~\ref{sec3}, we fit the explosion
energy evolution of multi-dimensional
simulations with our model.  To do so, we may choose three parameters
that are not entirely degenerate; we fit for the gain mass, $M_g$, the
starting shock radius, $r_0$, and the heating parameter, $\eta$.

\subsection{Analytic and Asymptotic Behavior of Explosion Energy }

In this section, we derive analytic solutions for the explosion
  energy for hypothesis I, including the asymptotic explosion energy, $E_{\infty}$.
  The evolution equation for the total dimensionless energy is 
\begin{equation}\label{dE_tilde_inf_dt_tilde}
\frac{d \tilde{E}}{d\tilde{t}}=\frac{\eta}{3(\gamma-1)\tilde{r}^2}+\frac{3\beta}{b\ln{(10)}} \frac{\tilde{\varv}}{\tilde{r}}\cosh^{-2}\left(\frac{11-\log_{10}(\rho_0/\tilde{r}^3)}{b}\right) \, .
\end{equation}
$\eta$ and $\beta$ are the constants of the model, so the path to analytic
  solution for $\tilde{E}$ is to find a simple
  analytic solution for $\tilde{r}$.  Since, $\tilde{\varv} = d
  \tilde{r}/dt$, our first challenge is to find an analytic solution
  for $\tilde{\varv}$.  To do this, we consider a Taylor series expansion
  about $\tilde{t} = 0$.  The initial conditions of this one-zone model
gives the first two terms: $\tilde{\varv}(\tilde{t} = 0) = 0$ and
$d\tilde{\varv}/dt(\tilde{t} = 0) = 0$.
The leading order term is the second derivative
with respect to $\tilde{t}$; taking the time derivative of
equation~(\ref{dvdttilde}) gives
\begin{equation}
\frac{d^2\tilde{\varv}}{d\tilde{t}^2}=\frac{2\tilde{\varv}}{\tilde{r}^3}+2\tilde{r}\tilde{\varv}\tilde{p}+\tilde{r}^2\frac{d\tilde{p}}{d\tilde{t}}=\eta
\, .
\end{equation}
The third order derivative is zero. \\
Up to leading order, the dimensionless velocity evolution is
\begin{equation} \label{Vassymptot}
\tilde{\varv}=\frac{\eta}{2}\tilde{t}^2 \, ,
\end{equation}
the evolution of the radius is
\begin{equation} \label{Rassymptot}
\tilde{r}=1+\frac{\eta \tilde{t}^3}{6} \, .
\end{equation}
Substituting this analytic expression for $\tilde{r}$ into
  equation~(\ref{dE_tilde_inf_dt_tilde}) and integrating over all time leads to an estimate for
  the asymptotic explosion energy:
\begin{equation} 
\begin{split}
\tilde{E}_\infty&=\frac{6^{1/3}\eta^{2/3}}{3(\gamma-1)}\int_0^\infty
\frac{d\tilde{t'}}{(1+\tilde{t}'^3)^2} \\
&+\frac{9\beta}{b\ln{(10)}}\int_0^\infty\frac{\tilde{t}'^2 d\tilde{t}'}{(1+\tilde{t}'^3)}\cosh^{-2}\left(\frac{11-\log_{10}{(\rho_0/(1+\tilde{t}'^3)^3)}}{b}\right)\, ,
\end{split}
\end{equation}
where $\tilde{t}'=(\frac{\eta}{6})^{1/3} \tilde{t}$. Integrating
  leads to a simple analytic expression
for the asymmptotic explosion energy:
\begin{equation}\label{Einf}
\tilde{E}_\infty\approx1.5\eta^{2/3}+\beta\left(1-\tanh\left(\frac{11-\log_{10}{(\rho_0)}}{b}\right)
\right) \, .
\end{equation}

\begin{figure}
\includegraphics[scale=0.5]{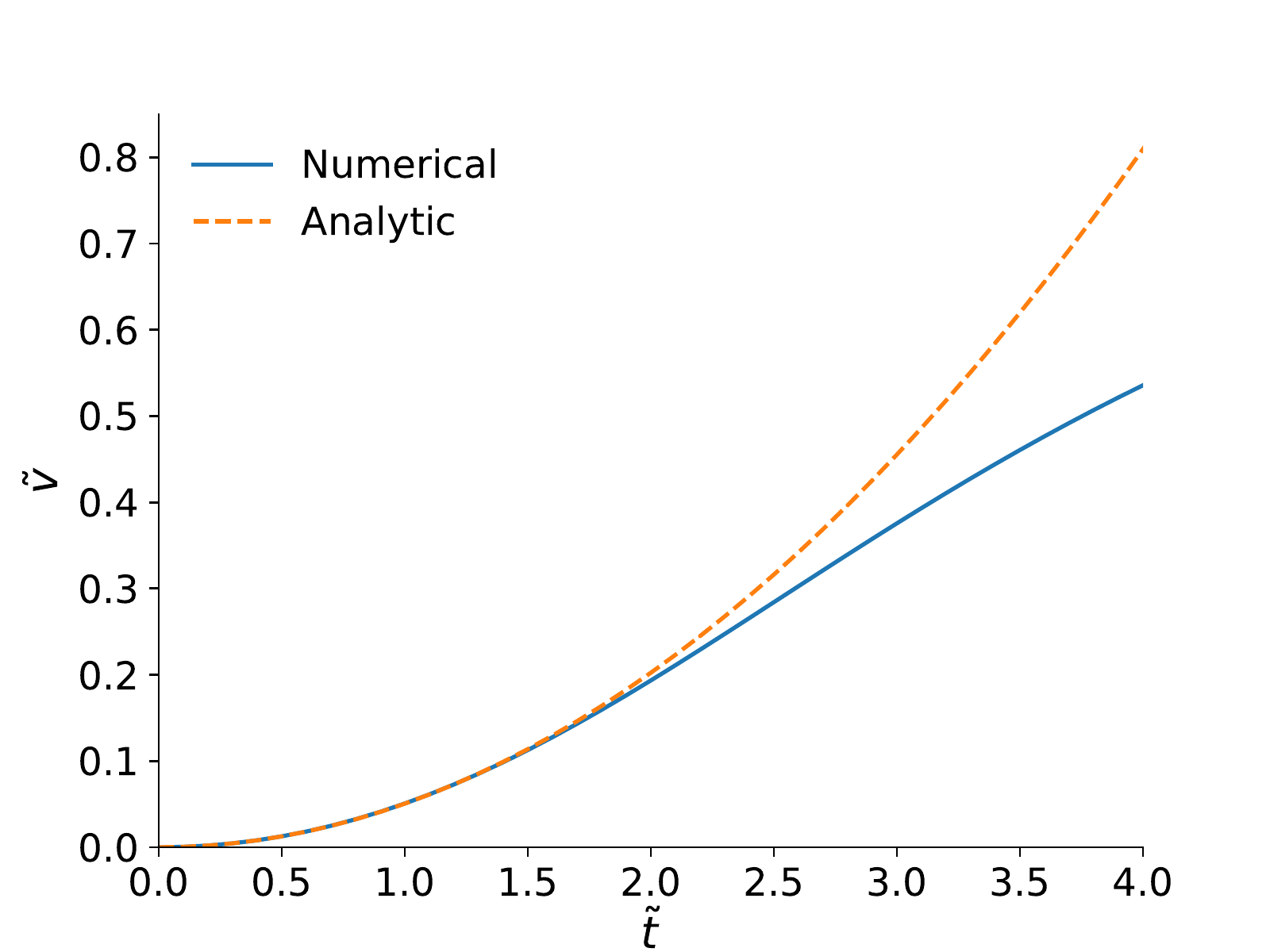}
\caption{Dimensionless velocity $\tilde{\varv}$ as a function of
  dimensionless time $\tilde{t}$ for 
$\eta=0.1$ and for hypothesis I. The solid line represents the
  numerical solution for velocity while the dashed line
  represents the quadratic approximation that we derive in
	equation~(\ref{Vassymptot}). For
  small values of $\tilde{t}$, the numerical and analytic expressions match.  }
\label{velocity}
\end{figure}

\begin{figure}
\includegraphics[scale=0.5]{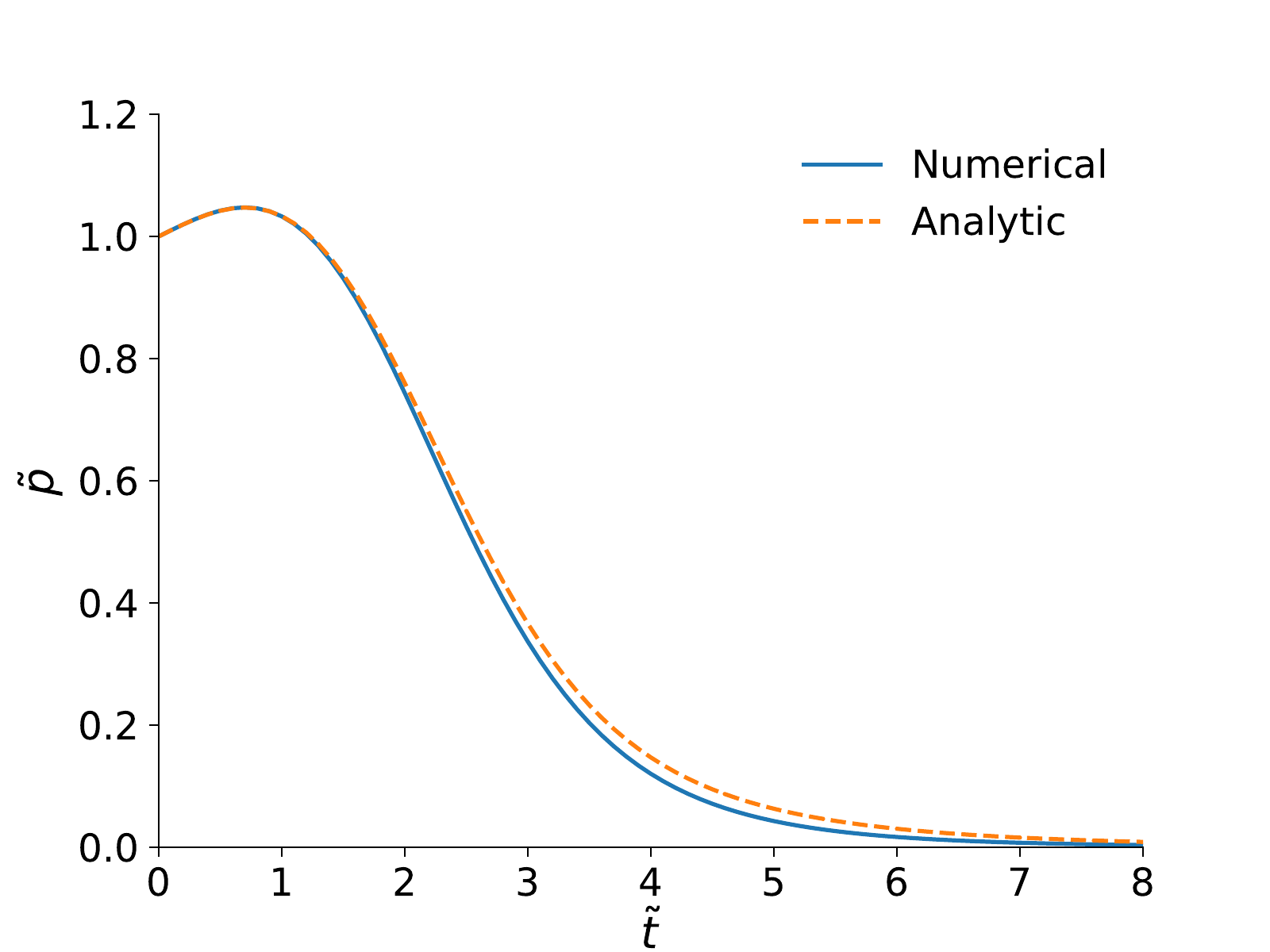}
\caption{Dimensionless pressure $\tilde{p}$ as a function of
  dimensionless time $\tilde{t}$ for $\eta=0.1$ and for
	hypothesis I. As in Figure~\ref{velocity}, the solid line
  represents the numerical solution for velocity, while
  the dashed line is the analytic solution of
  equation~(\ref{dvdttilde}). The analytic approximations produce excellent
	agreement even at late times.  
\label{analyticpressure}}
\end{figure}

\begin{figure}
\includegraphics[scale=0.5]{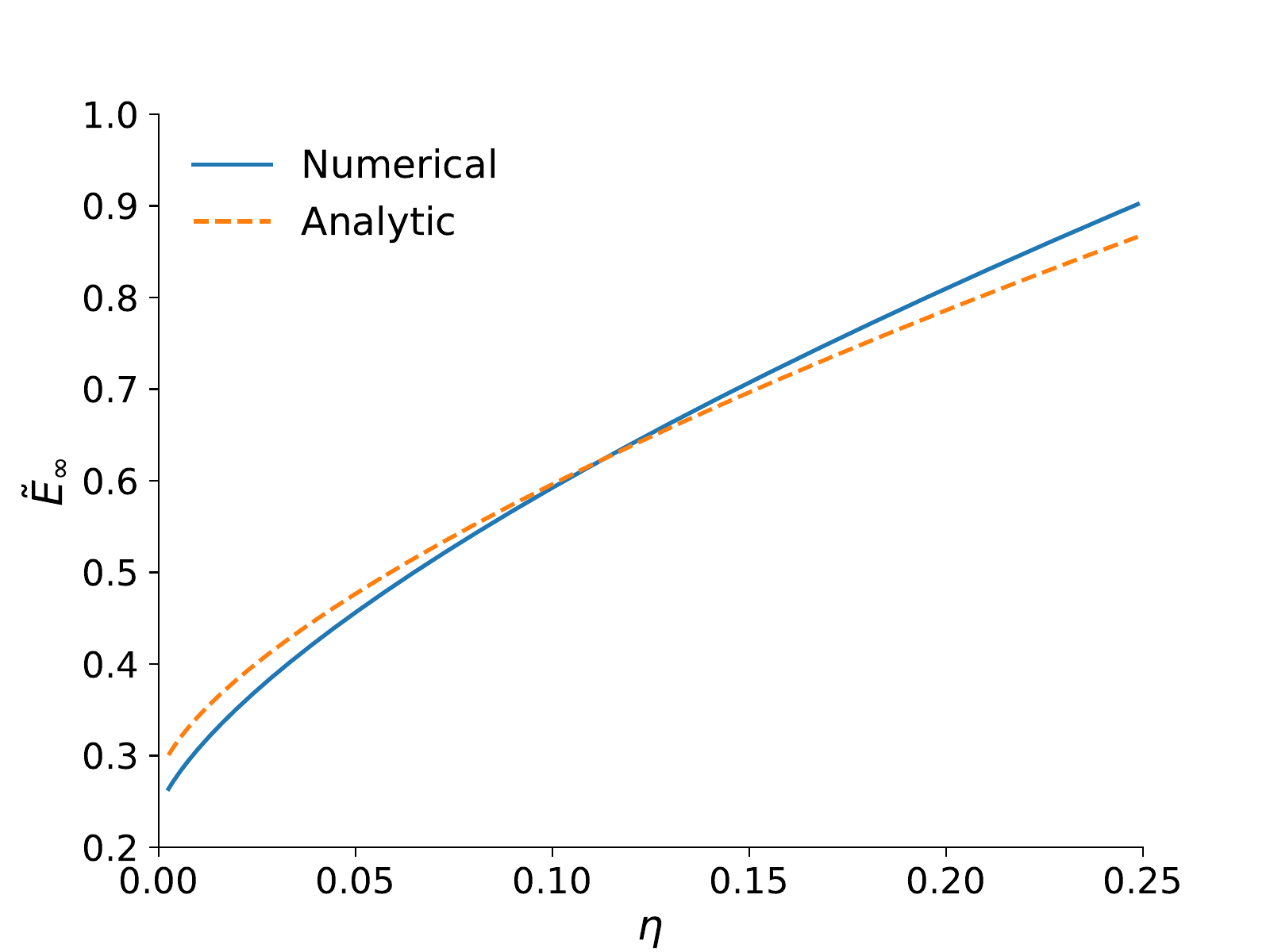}
\caption{$\tilde{E}_\infty$ as a function of $\eta$ for
	hypothesis I.  As in
	Figures~\ref{velocity}~\&~\ref{analyticpressure}, the solid line represents the numerical
	result, and the dashed line represents our analytic result,
	equation~\ref{Einf}.  We find that  the analytic expression matches with numerical
  result  with less than a 15\% discrepancy.}
\label{EvsEta}
\end{figure}

Figures~\ref{velocity}-\ref{EvsEta} compare these analytic
  solutions with the numerical solution of equations~(\ref{dvdttilde})-~(\ref{dpdttilde}).  First,
  Figure~\ref{velocity} compares the analytic solution for $\tilde{\varv}$
  with the numerical solution.  As derived, the leading order is the
  second order term, and the value of the derivative is the heating
  parameter, $\eta$.  For times later than $\tilde{t} \approx 2$, the
  analytic and numerical solutions begin to diverge.  However, because
  $\tilde{r}^2$ appears in the demoninator of the evolution equation
  for $\tilde{E}$, equation~(\ref{dE_tilde_inf_dt_tilde}), the evolution of $\tilde{E}$ is
  most sensitive to the early time evolution of $\tilde{\varv}$ and
  $\tilde{r}$.  Hence, the divergence between the analytic and
  numerical solutions at late times do not present a problem.

Figure~\ref{analyticpressure} compares the analytic pressure with
the numerical solution.  To find the analytic pressure, we take the
first derivative of the analytic velocity and use equation~(\ref{dvdttilde}) to
find a solution for $\tilde{p}$.  In an absolute sense, the analytic
and numerical pressures agree for all time.  This agreement is due to
the radius appearing in the denominator of the integrand in
equation~(\ref{dvdttilde}).  In a relative sense, the analytic and numerical
solutions do diverge at late times in a similar way as for the velocity.

Figure~\ref{EvsEta} compares the analytic and numerical solutions
  for the asymptotic explosion energy as a function of $\eta$.  The
  analytic result that $\tilde{E}_{\infty} \approx 1.5 \eta^{2/3}+\beta\left(1-\tanh\left(\frac{11-\log_{10}{(\rho_0)}}{b}\right)\right)$
  matches the numerical explosion within an accuracy of 15\%. 
  
In summary, the dimensional explosion energy
for hypothesis I is $\tilde{E}_\infty \approx 1.5\eta^{2/3}+\beta\left(1-\tanh\left(\frac{11-log_{10}(\rho_0)}{b}\right)\right)$ and for hypothesis II is $\tilde{E}_\infty \approx 
1.5 \eta^{2/3}$.  The total explosion energy is 
\begin{equation}
E_\infty \approx M_g v_0^2
\left[1.5\eta^{2/3}+\beta\left(1-\tanh\left(\frac{11-log_{10}(\rho_0)}{b}\right)\right)\right].
\label{E_tot}
\end{equation}
 We find that for reasonable values of
  $\eta$, the recombination energy contributes a significant
  fraction of the explosion energy.
  For example, for $\eta=0.15$, $M_g=0.05 M_{\odot}$, and $r_0=150 km$, $\approx 50$\% of the
  explosion energy is determined by $\alpha$ recombination.

\section{Comparing Analytic Theory with Multi-dimensional Simulations} \label{sec3}

As a preliminary test of the simple explosion model, we
  use Bayesian statistical inference to fit the model to
  multi-dimensional simulations. The simulations originate from four
  codes (CHIMERA, CoCoNuT-FMT, FORNAX2D, FORNAX3D), and a description of all four codes and their
  capabilities is summarized in \citet{murphy2019}.  In essence, the
  explosion model is a theoretical prediction and we use the
  multi-dimensional simulations to test this prediction.  The model
has a characteristic curve defined by three parameters.  This
characteristic curve may or may not fit the explosion energy curves of
the multi-dimensional simulations.  If they do not fit, then this test
would rule out the simple model.  Even if the curve does fit, the
model parameters may not match typical scenarios of core-collapse
simulations.  Our goal is
  to test the predictions of the simple model in two ways.  For one, we will
  test if the model provides a good fit to the simulations.  Second,
  if the model produces a good fit, we will assess whether the model
  parameters are reasonable.

With these two tests, broadly, there are three possible
  outcomes.  For one, the model may produce a reasonable fit and
  parameters.  This scenario would suggest that the model is a good
  theoretical description of the explosion evolution.  Since the simple model is inherently
spherical, we expect that explosions of low mass progenitors are the
most likely to be consistent with the model.  On the other hand, more
massive progenitors tend to explode aspherically, so we do not expect
the simple model to fit these inherently multi-dimensional
explosions.

A second scenario is that the simple model fits the
  multi-dimensional results but the model parameters do not match
  typical simulation conditions.  This would not completely rule out
  the model, but it would certainly call into question the fidelity of
the model in representing the simulations.

The third scenario is that the model neither produces a
  reasonable fit nor reasonable parameters.  This would rule out the
  simple model as a good theory.  The obviously multi-dimensional
  explosions of the most massive stars would likely fall into this
  category.  This would also suggest (but not prove) that the
  development of the explosion dynamics are indeed multi-dimensional.

\subsection{Bayesian Fitting}

We use Bayesian statistical inference to both provide model fits
  and to infer the model parameters.  Generically,
  Bayesian inference provides a way to statistically infer theory
  parameters given data.  In this paper, the simple explosion model
	represents the theory, and the explosion energy evolution of the
	multi-dimensional simulation represents the data.
The model parameters are $\theta = \{ M_g, r_0,
  \eta\}$, 
    and $y$ is the explosion energy as a function of time for a
	  single simulation.  

The posterior distribution of the model
  parameters, $\theta$ given simulation explosion
  energy evolution, $y$ is 
\begin{equation}
p(\theta|y)=\frac{p(\theta) \mathcal{L}(y|\theta)}{p(y)} \, .
\end{equation}
$p(\theta)$ is the prior distribution for the model parameters; for
this initial analysis we assume uninformative (uniform) priors for all
parameters.  $\mathcal{L}(y|\theta)$ is the likelihood or sampling distribution, $p(y)=\sum_\theta p(\theta)p(y|\theta)$ is the  normilization.

To construct the likelihood, we first consider the likelihood at
  each time, indexed by $k$: 
\begin{equation}
\mathcal{L}_k (y|\theta)=\frac{1}{\sqrt{2 \pi \epsilon^2}}
e^{-\frac{(E_{\text{sim},k}-E_k)^2}{2\epsilon^2}} \, .
\end{equation}
$E_{\text{sim},k}$ is the explosion energy at time index $k$, and
$E_k$ is the simple model explosion energy at the same time.
 $\epsilon$ represents scatter in the simulation,
  which might be due to inherent scatter in the underlying simulation
  or measurement of the explosion energy.  In either case, this
  scatter is unknown and is a nusiance parameter.   The total likelihood is
\begin{equation}
\mathcal{L}(y|\theta)=\prod_k \mathcal{L}_k(y|\theta) \, .
\end{equation} 

To infer the  posterior distribution, we use a Markov Chain Monte
Carlo (MCMC) package called EMCEE (\cite{Foreman_Mackey_2013},
\cite{goodman2010}). EMCEE constructs posterior distribution by
finding Likelihood function on each step of the calculation. This
process requires information about explosion energy as a function of
our set of parameters $\theta$ and is time-sensitive. We avoid solving
our differential equations for each step in the chain by providing
dimensionless energy $\tilde{E}(\tilde{t},\eta,\beta)$ as a look-up table. Afterward, we
interpolate the look-up table by using tri-linear interpolation.
Again, our independent parameters are $M_g$,$r_0$ and $\eta$. To
restrict the parameters to physically valid values, we set the bounds
of the uniform priors to be $M_g < 0.5 M_{\odot}$ and $\eta<0.25$ (corresponding to neutrino luminosity $L<10^{53} erg/s$).
For the MCMC runs, we typically use 100 walkers, 4000 steps and we
burn 1000 of those.

\subsubsection{Bayesian Fitting Results}

Using Bayesian inference, we fit our theoretical model to
multi-dimensional simulations. As expected, the theory fits those
  simulations that have mostly spherically symmetric explosions and
  does not fit the predominantly aspherical explosions.

\begin{figure} 
\includegraphics[scale=0.5]{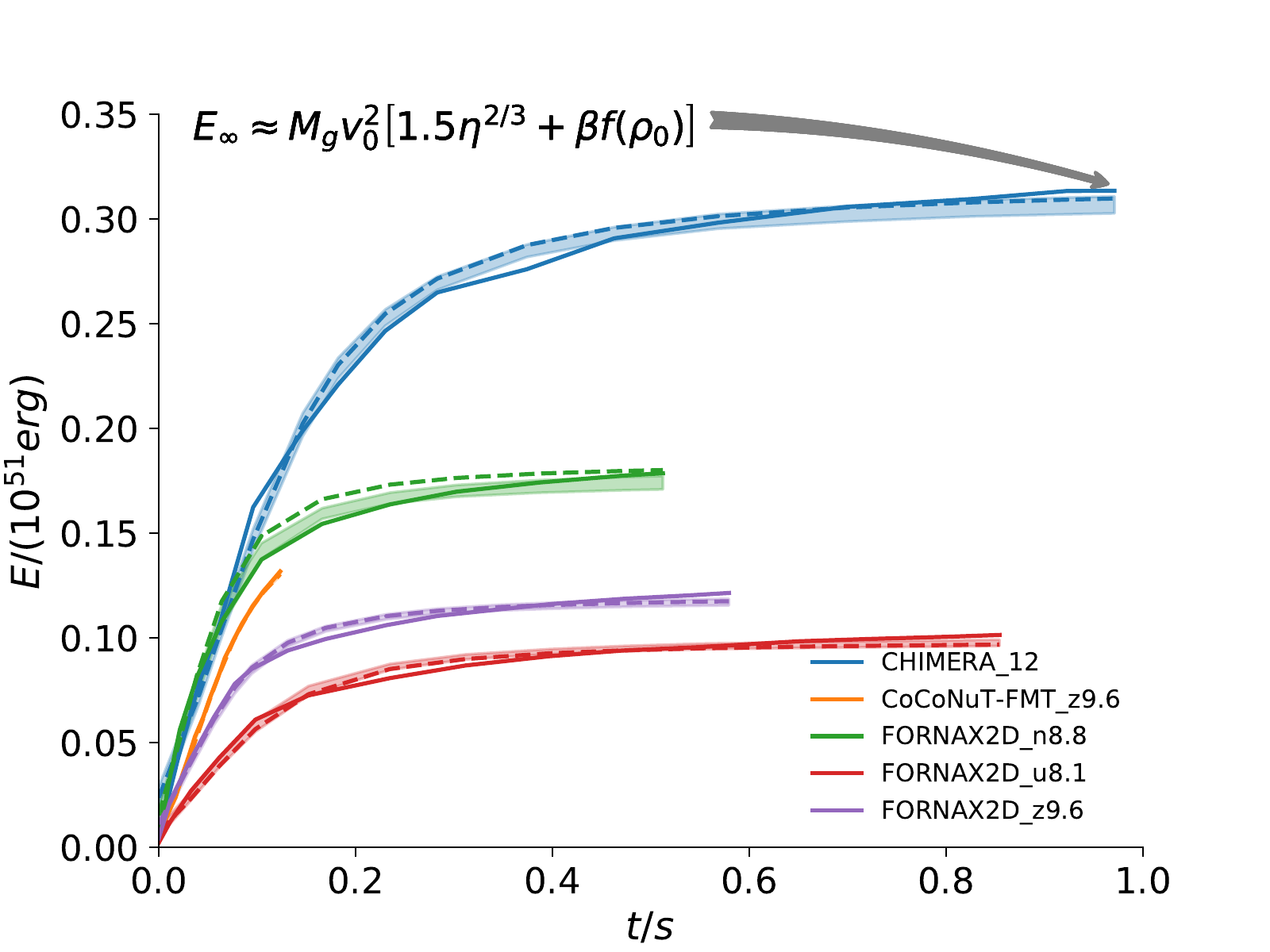}
\includegraphics[scale=0.5]{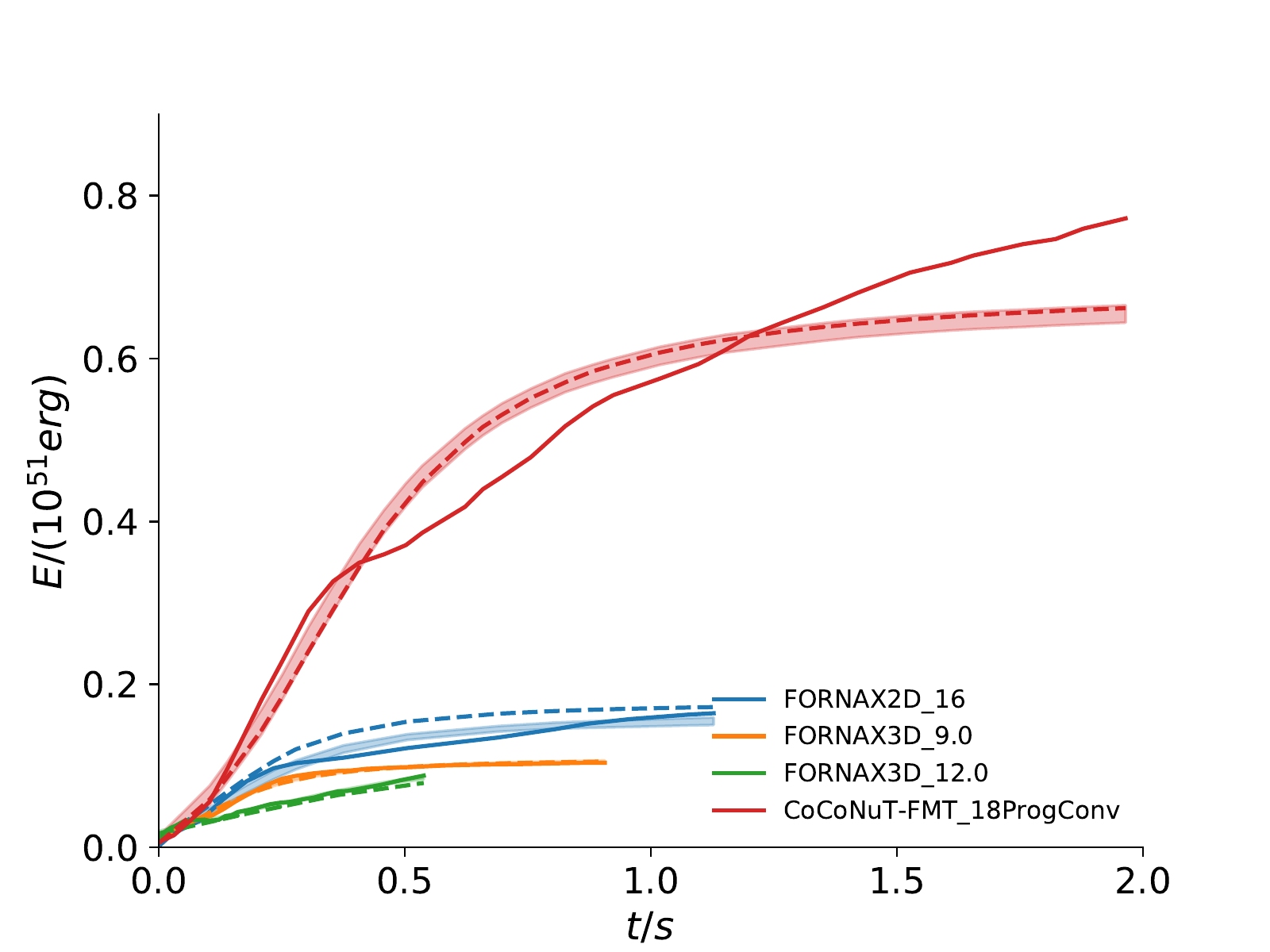}
\caption{Explosion energy vs time after bounce for hypothesis I
  (neutrino power + $\alpha$ recombination). The solid
  lines represent the simulation curves, while
  the shaded bands represent fits of the theoretical model.The dashed
  line represents energy curve for the best fit.  The upper panel
  shows the five models with good fits and reasonable parameters;
	the bottom panel shows some of the poor fits. In general,
	the good fits represent simulations of low progenitor masses,
	which explode mostly spherically.  The poor fits represent higher
	progenitor mass simulations, which tend to explode aspherically. The analytic explosion model is consistent with the
	mostly spherically symmetric explosion, and, not surprisingly, it
	is not consistent with the aspherical explosions.  This suggests
	that aspherical dynamics are important in determining eventual
	explosion energies for the higher progenitor mass simulations.
 }
\label{rec}
\end{figure}

Figure \ref{rec} shows the  Bayesian fits for hypothesis I (neutrino power + $\alpha$
  recombination). The upper panel represents a sample of the
good fits, and the lower panel shows some of the poor
fits. Generally, the simulations in the top
panel explode spherically and simulate the explosion of lower
  mass progenitors, while
simulations in the lower panel explode
aspherically and represent more massive progenitors. The
explosion model is consistent with the spherical explosions, but
  is inconsistent with the aspherical explosions.  This suggests that
 aspherical dynamics may be important in the explosion evolution of
 the aspherical explosions.

\begin{figure*}
\includegraphics[scale=0.3]{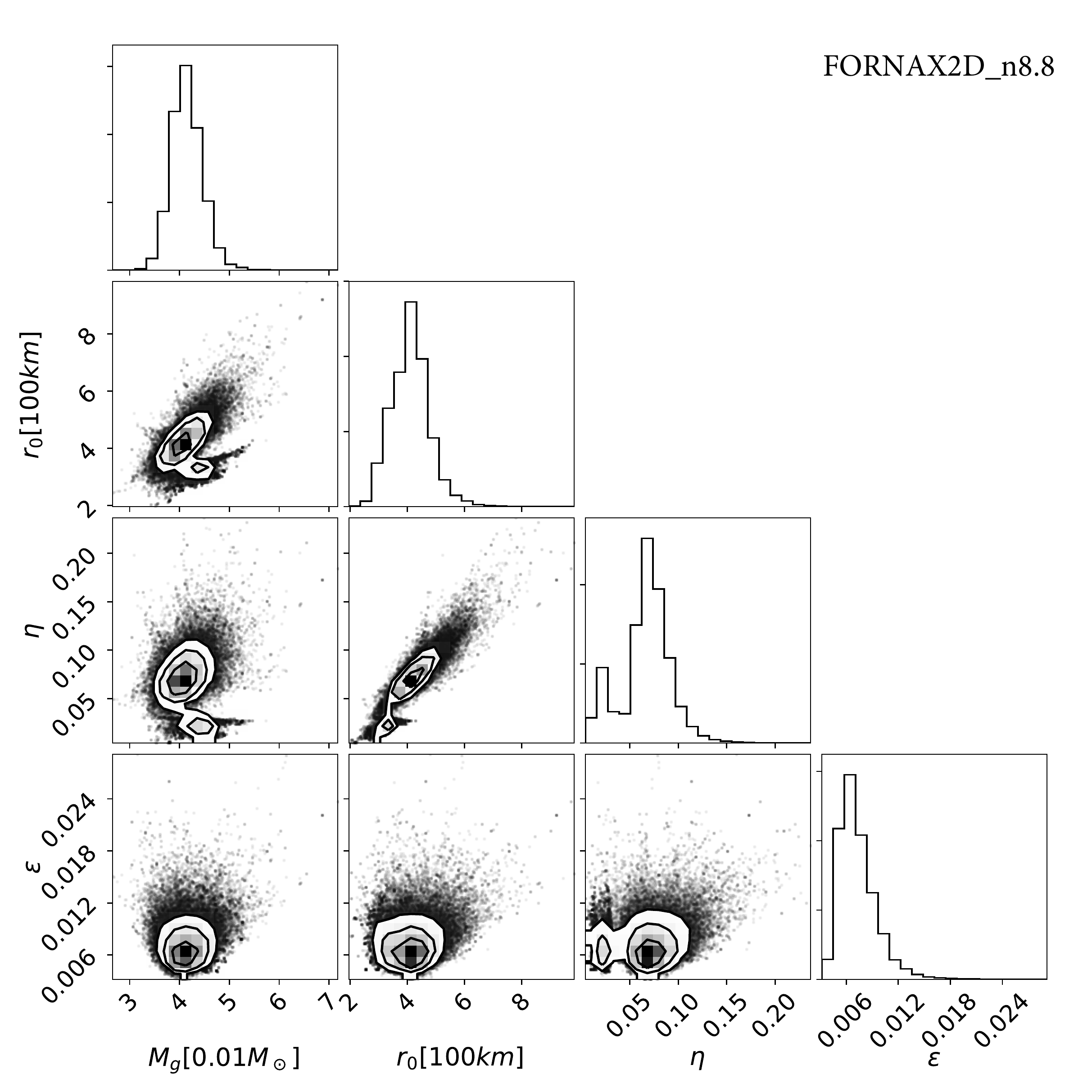} 
\includegraphics[scale=0.3]{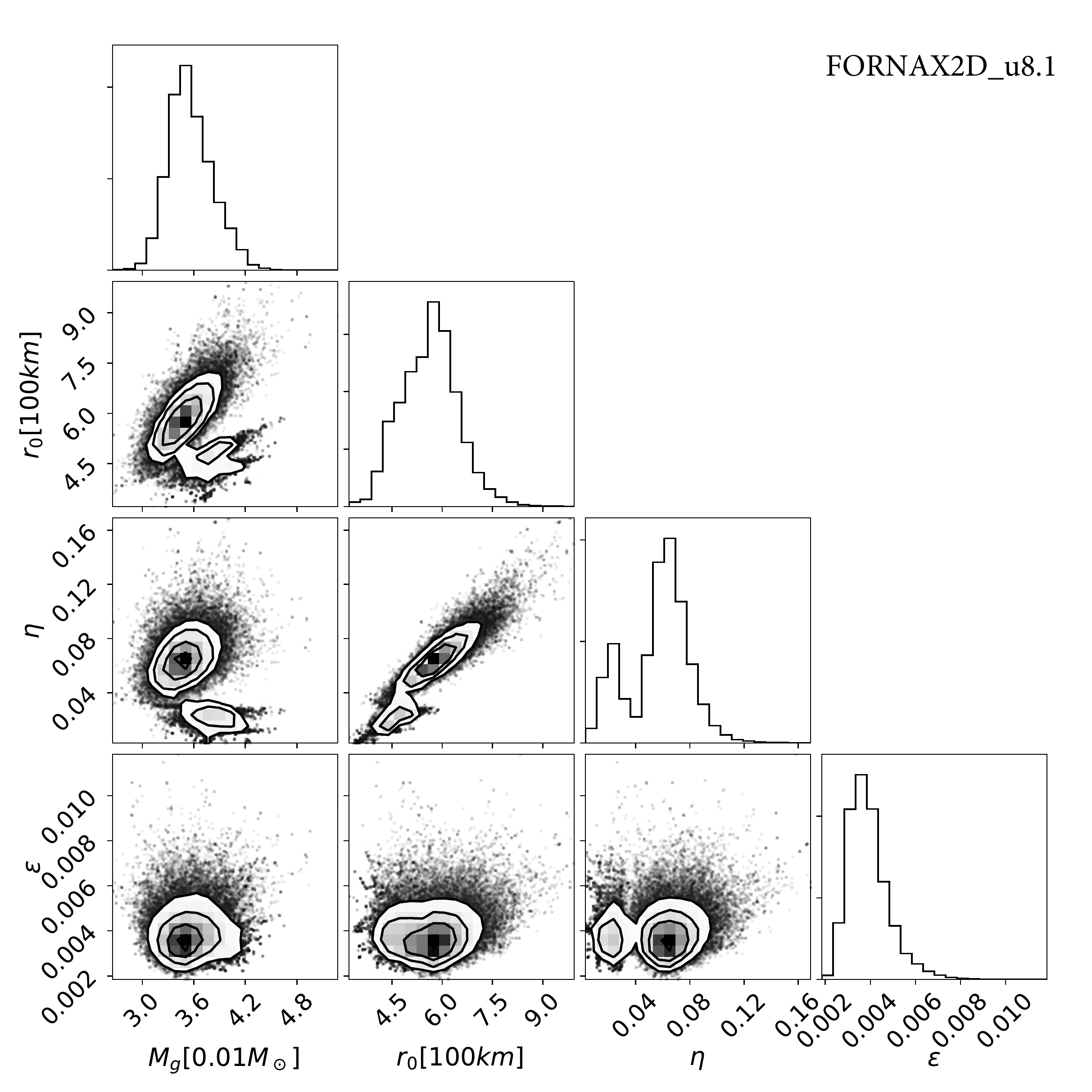}
\includegraphics[scale=0.3]{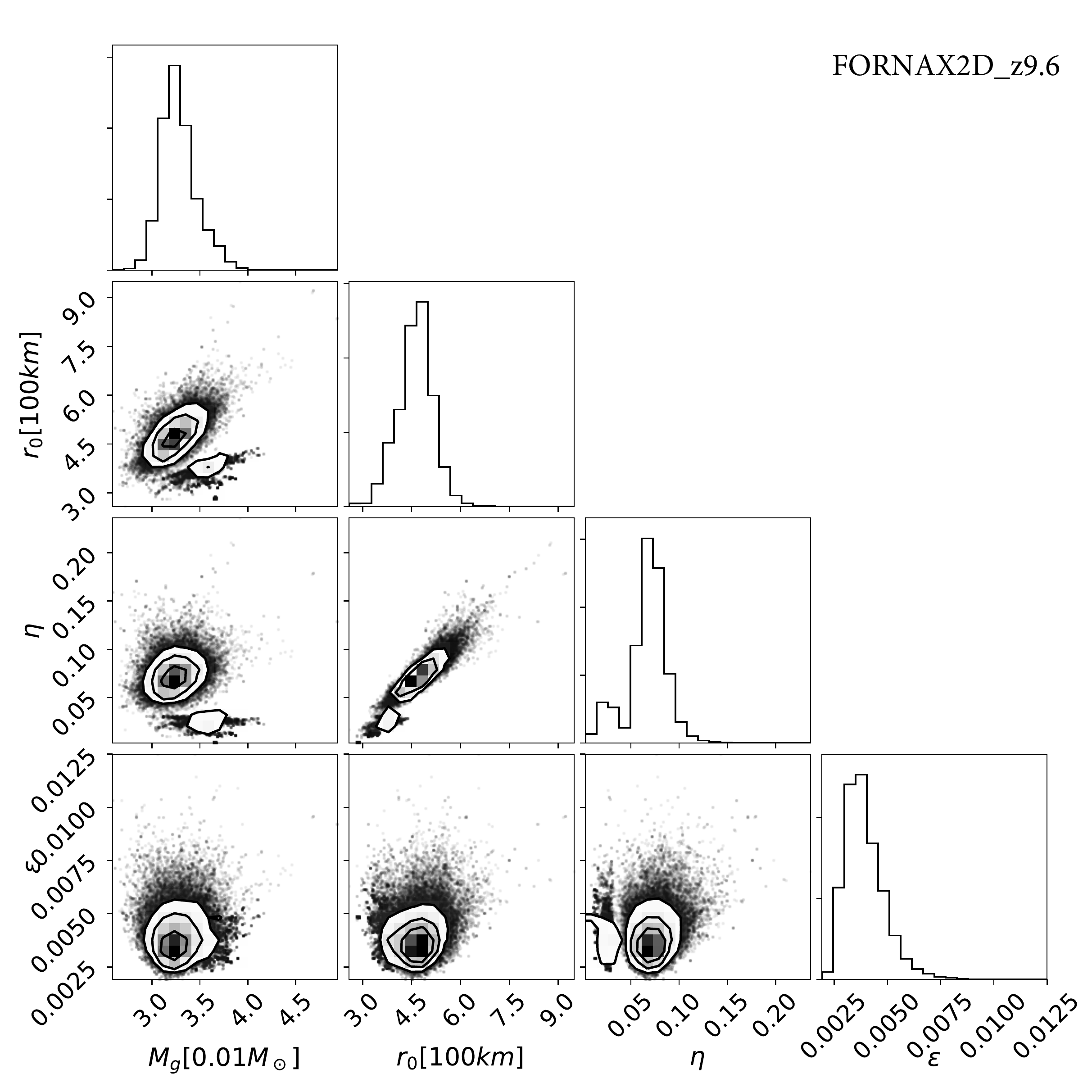}
\includegraphics[scale=0.3]{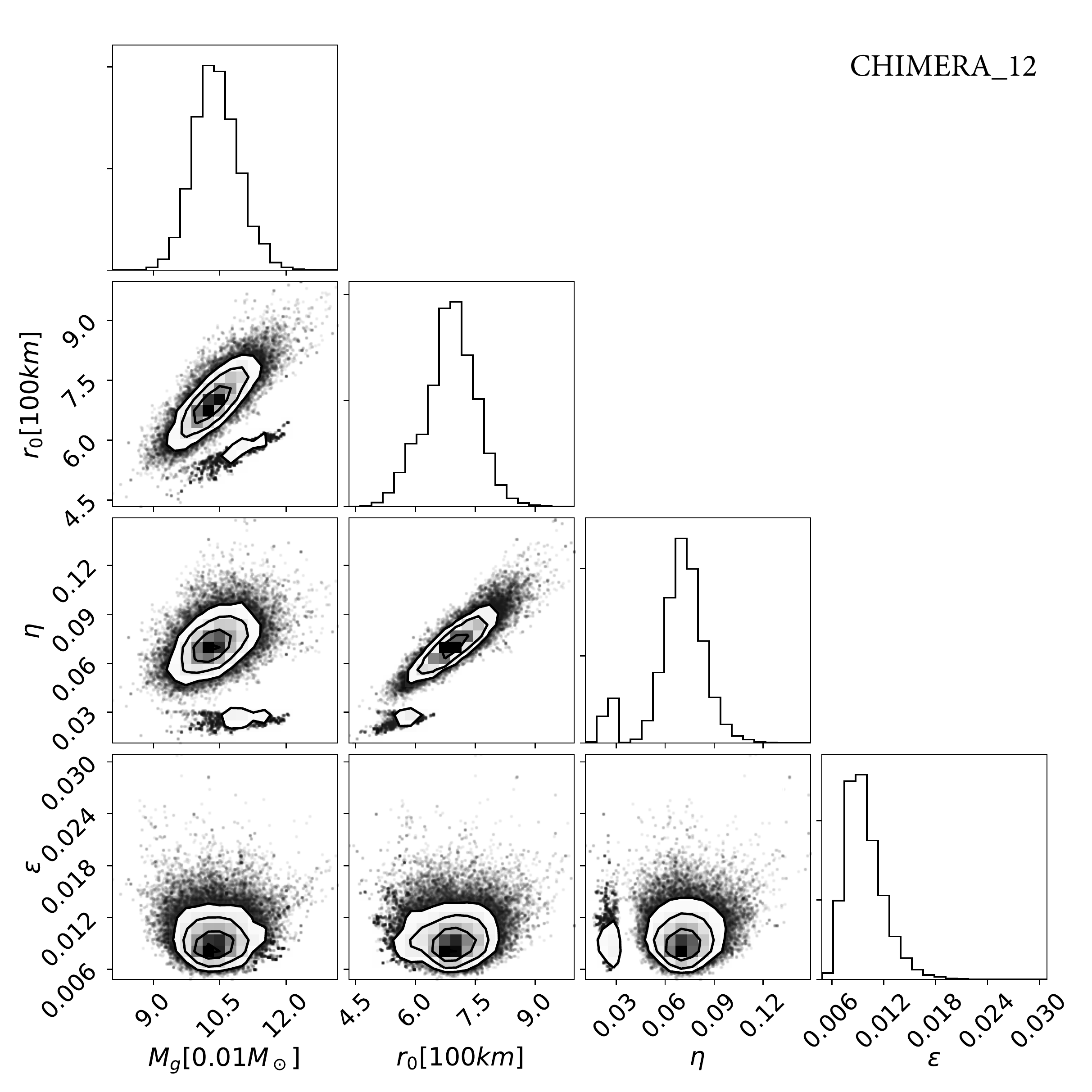}
\includegraphics[scale=0.3]{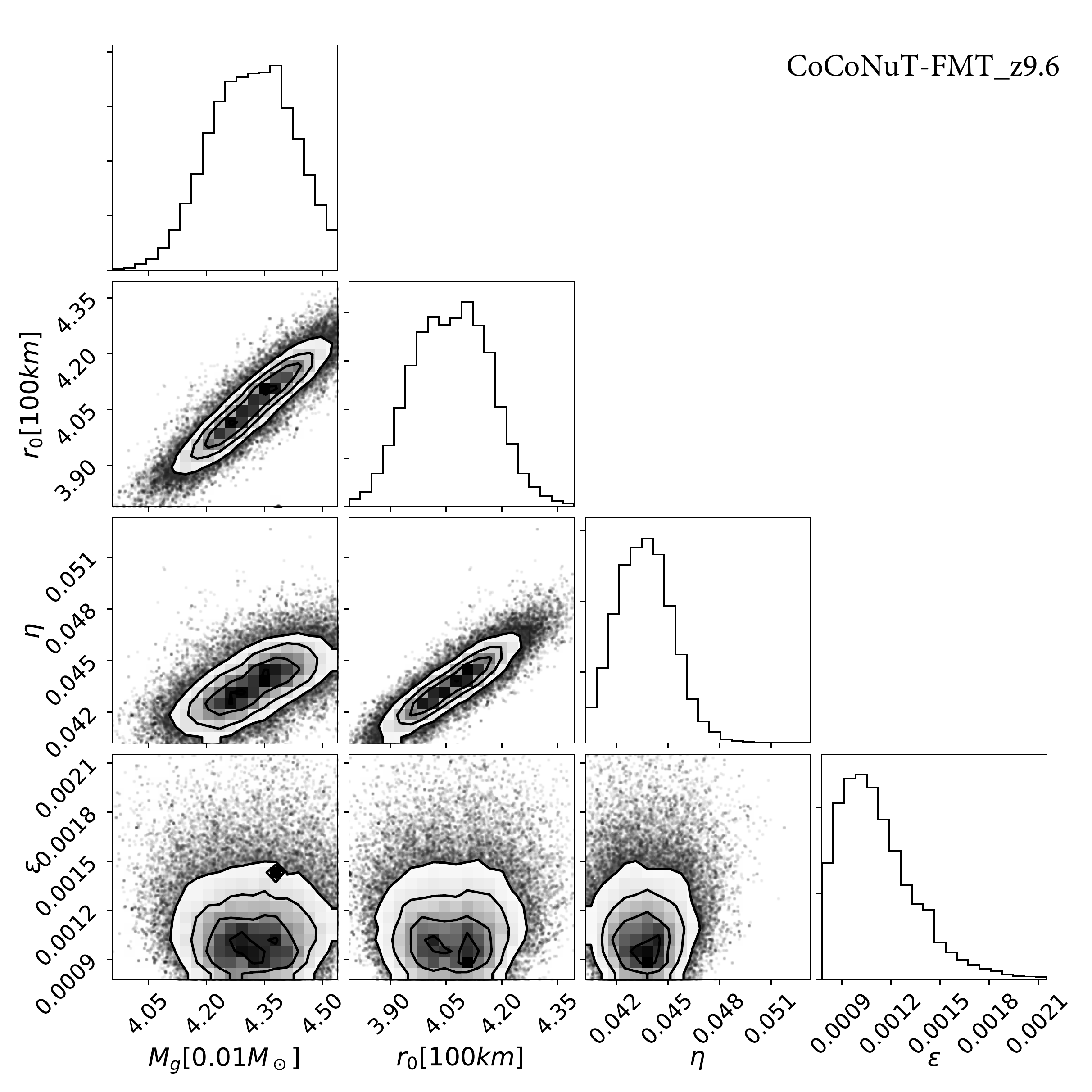}

\caption{The posterior distribution for the explosion model
	parameters, $M_g$ ,$r_0$, $\eta$ , and $\epsilon$ for hypothesis I.
  This fit corresponds
   to Fornax2D n8.8 \citep{Radice2017},Fornax2D u8.1 \citep{Radice2017}, Fornax2D z9.6 \citep{Radice2017},Chimera 12 \citep{Bruenn2016},CoCoNuT-FMT z9.6 \citep{Muller2019} .  There is a degeneracy
   among the parameters given by eq.~(\ref{eta}). }
\label{8.8}
\end{figure*}

Figure~\ref{8.8} 
shows the posterior distributions for the good fits (top panel of
Figure~\ref{rec}). The best-fit model parameters for each simulation
are as follows: 
FORNAX2D n8.8 --- $M_g=0.0413^{+0.0032}_{-0.0027}$ M$_\odot$, $r_0=
411^{+60}_{-68}$ km, $\eta=0.07^{+0.02}_{-0.04}$; 
FORNAX2D u8.1 --- $M_g=0.0353^{+0.0029}_{-0.0021}$ M$_\odot$, $r_0=
567^{+74}_{-93}$ km, $\eta=0.06^{+0.02}_{-0.03}$;
FORNAX2D z9.6 --- $M_g=0.0326^{+0.0020}_{-0.0015}$ M$_\odot$, $r_0=
465^{+46}_{-59}$ km, $\eta=0.07^{+0.01}_{-0.02}$; 
CHIMERA 12 --- $M_g=0.1039^{+0.0052}_{-0.0046}$ M$_\odot$, $r_0=
690^{+59}_{-66}$ km, $\eta=0.07^{+0.01}_{-0.01}$; 
CoCoNuT-FMT z9.6 --- $M_g=0.0432^{+0.0012}_{-0.0012}$ M$_\odot$, $r_0=
407^{+11}_{-11}$ km, $\eta=0.04^{+0.00}_{-0.00}$.
The uncertainties represent the 68\% highest density confidence
intervals.  In addition to providing a good
  fit, these parameters are also roughly consistent with values
  measured in CCSN simulations.  A detailed comparison with the
  simulations could help to further constrain the theoretical model.

\begin{figure}
\includegraphics[scale=0.5]{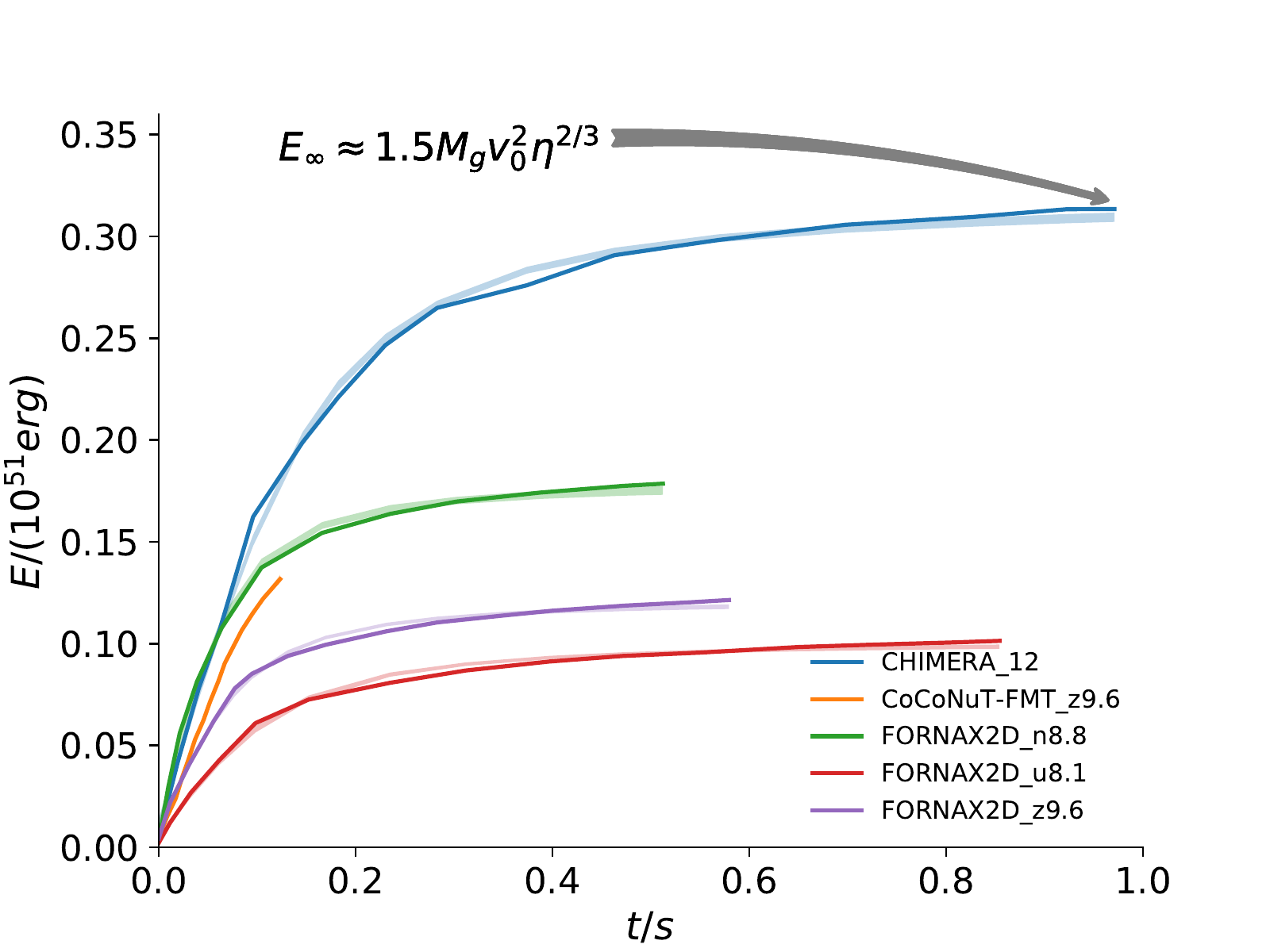}
\includegraphics[scale=0.5]{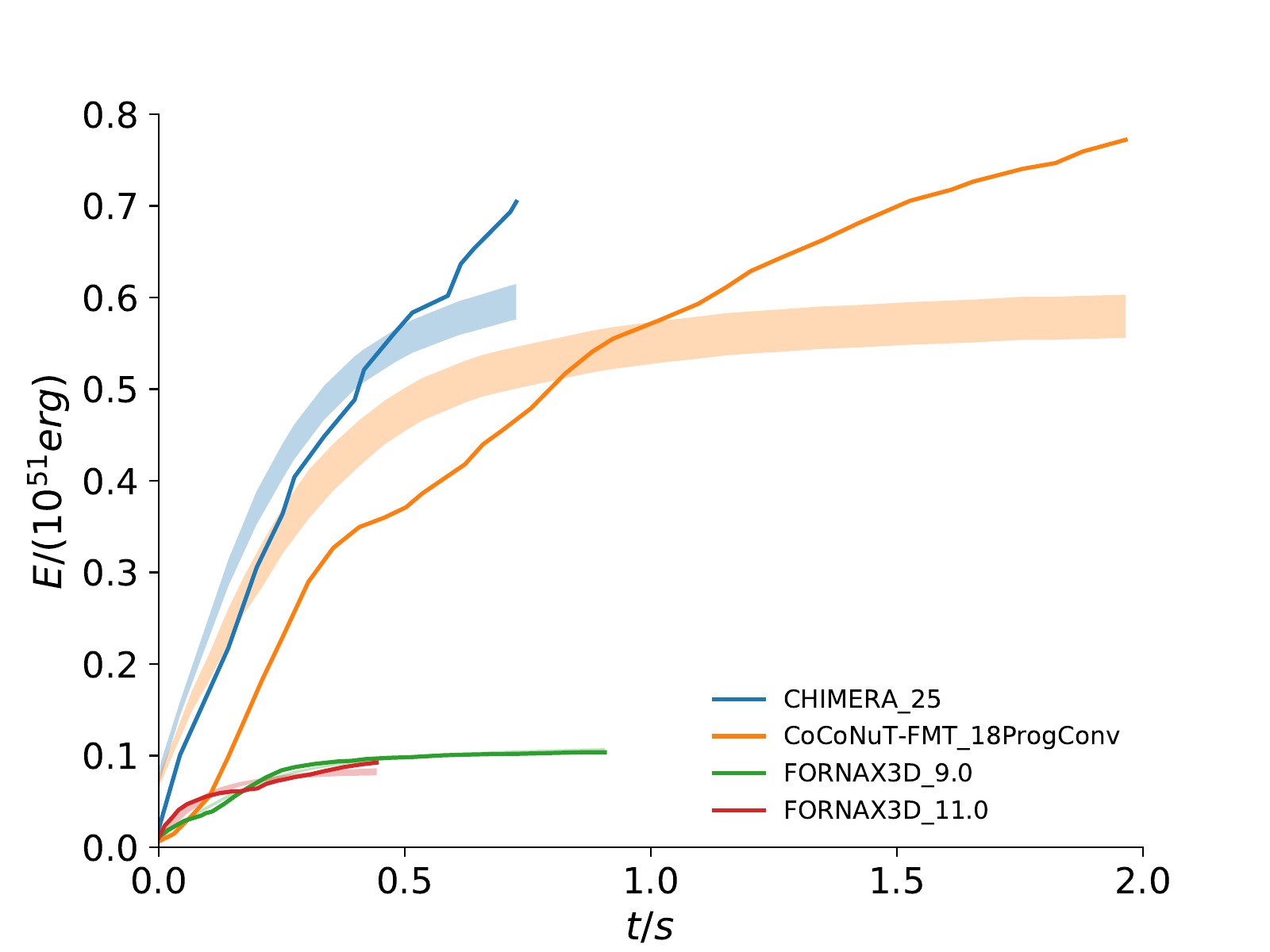}
\caption{Similar to Figure~\ref{rec} except for hypothesis II.  As is the case for hypothesis I, the
	explosion model of hypothesis II is consistent with the
	mostly spherically symmetric explosion and not consistent with the aspherical explosions.  Again, this suggests
	that aspherical dynamics are important in determining eventual
	explosion energies of the higher progenitor mass simulations.}
 \label{EvsT}
\end{figure}

Figure \ref{EvsT}  shows the Bayesian fits for hypothesis II (neutrino power only). As
  in Figure~\ref{rec}, the upper panel presents 
good fits, while the lower panel shows some of the poor
fits. Again, the simulations on the top panel explode spherically, while
 those in the lower panel explode aspherically. The
fact that our spherically symmetric model fits the simulations given
on the top panel and cannot fit lower ones is consistent with the
  hypothesis that some simulations explode spherically while others
  are dominated by aspherical dynamics.

\subsection{Goodness-of-fit with Predictive Posterior Checking}

An MCMC method will find the most likely model parameters, but it
  does not determine whether the theory and model are a good fit to
  the simulation data.  Therefore, we use posterior predictive
  checking to check for goodness-of-fit.  The goal of this technique
  is to compare the predicted ``data'' from the simple model with the
``data'' from the simulations.  If the model produces a good fit, then
the predicted ``data'' should match the simulation ``data''.

As before, $y$ represents the ``data'' or explosion energy as a
  function of time from the simulations.  $y^{\text{rep}}$ represents
  the replicated ``data'' from the model, and $\theta$ are the model
  parameters.  In the previous section, we describe how to infer the
  model parameters, $\theta$, given the data, $y$.  One then may use
  those model parameters to calculate the model explosion energy
  curve; this is the replicated data, $y^{\text{rep}}$, given the original
  data, $y$, or explosion energy curve from the simulation.  The goal
  of our goodness-of-fit test is to compare the replicated data with the
  original data.

The following describes the specific goodness-of-fit test that we
  use.  First of all, note that there are $N_s$ replicated data for
  each time, $t_k$, where $N_s$ is the number of MCMC samples.  If the replicated data is a good representation
  (i.e. good fit) of the data, then half of these replicated data
  should be above the data at time $t_k$ and the other half should be
  below.  This is the essence of our goodness-of-fit test.  The first
  step is to calculate the probability that $y^{\text{rep}}_k \leqslant y_k$:
\begin{equation}
P_{\leqslant} = P_k \left(y^{\text{rep}}_k \leqslant y_k|y \right) = \int
P(y^{\text{rep}}_k \leqslant y_k | \theta) P(\theta| y) d\theta \, .
\end{equation}
In practice the right-hand is discrete.  At each time $t_k$, we
calculate replicated data by taking the $N_s$ MCMC samples of
$\theta$ as inputs to our model.  This generates $N_s$ replicated data
at each time $t_k$.  We then calculate the fraction of these
replicated data that are less than or equal to the data (explosion energy
from simulation).

If the replicated data accurately represents the data, then $P_{\leqslant}$
  should be distributed as a uniform distribution between 0 and 1:
\begin{equation}
P(P_{\leqslant}) \thicksim \mathcal{U}(0,1) \, ,
\end{equation}
where $P(P_{\leqslant})$ is the distribution of $P_{\leqslant}$ for all times, and $\mathcal{U}(0,1)$ is the uniform distribution.  To test
  whether $P(P_{\leqslant})$ is indeed drawn from the uniform distribution, we
  use the one-sided Kolmogorov-Smirnov test.

\subsection{Goodness-of-fit Results}

\begin{figure}
\includegraphics[scale=0.5]{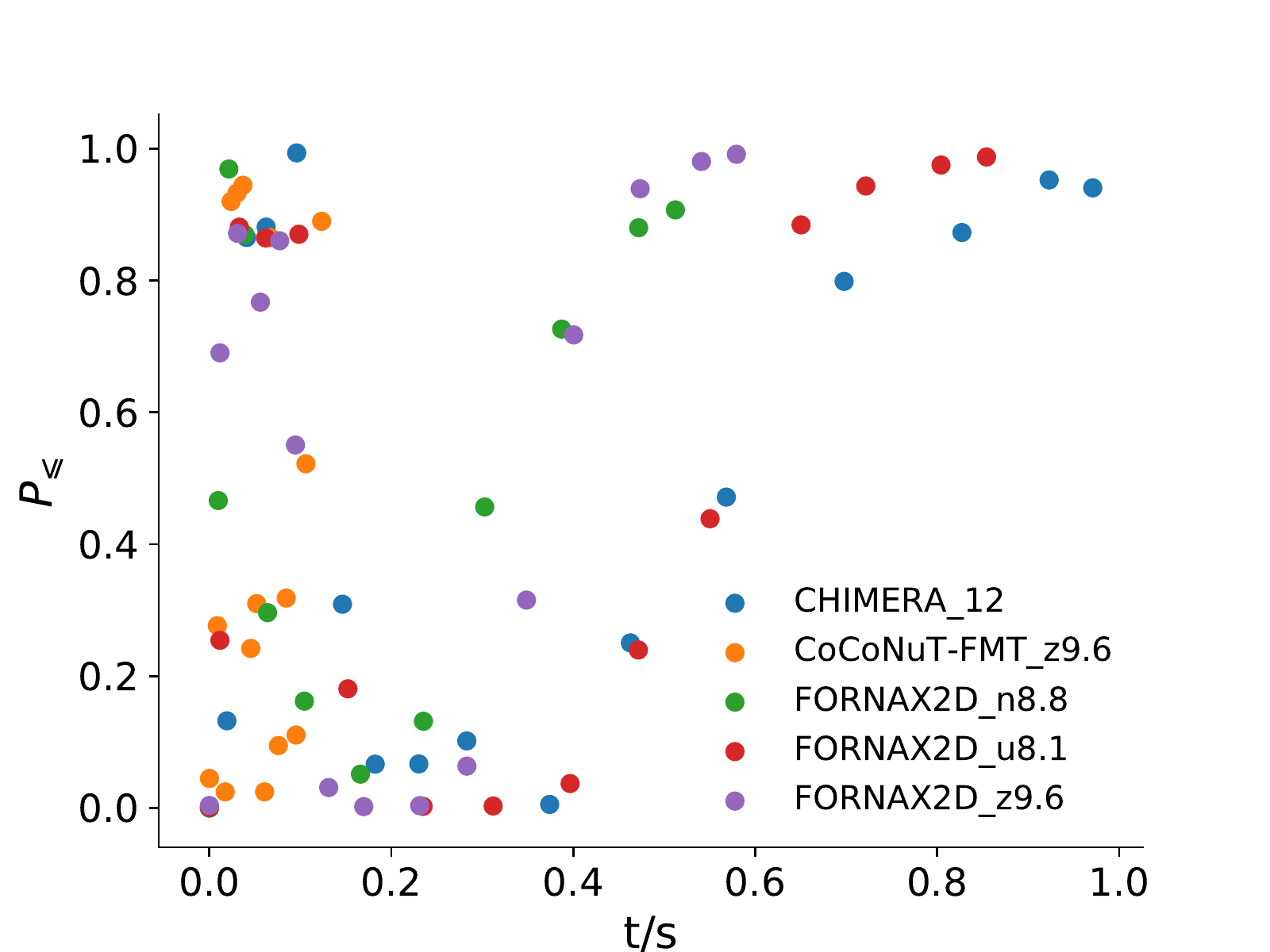}
\includegraphics[scale=0.5]{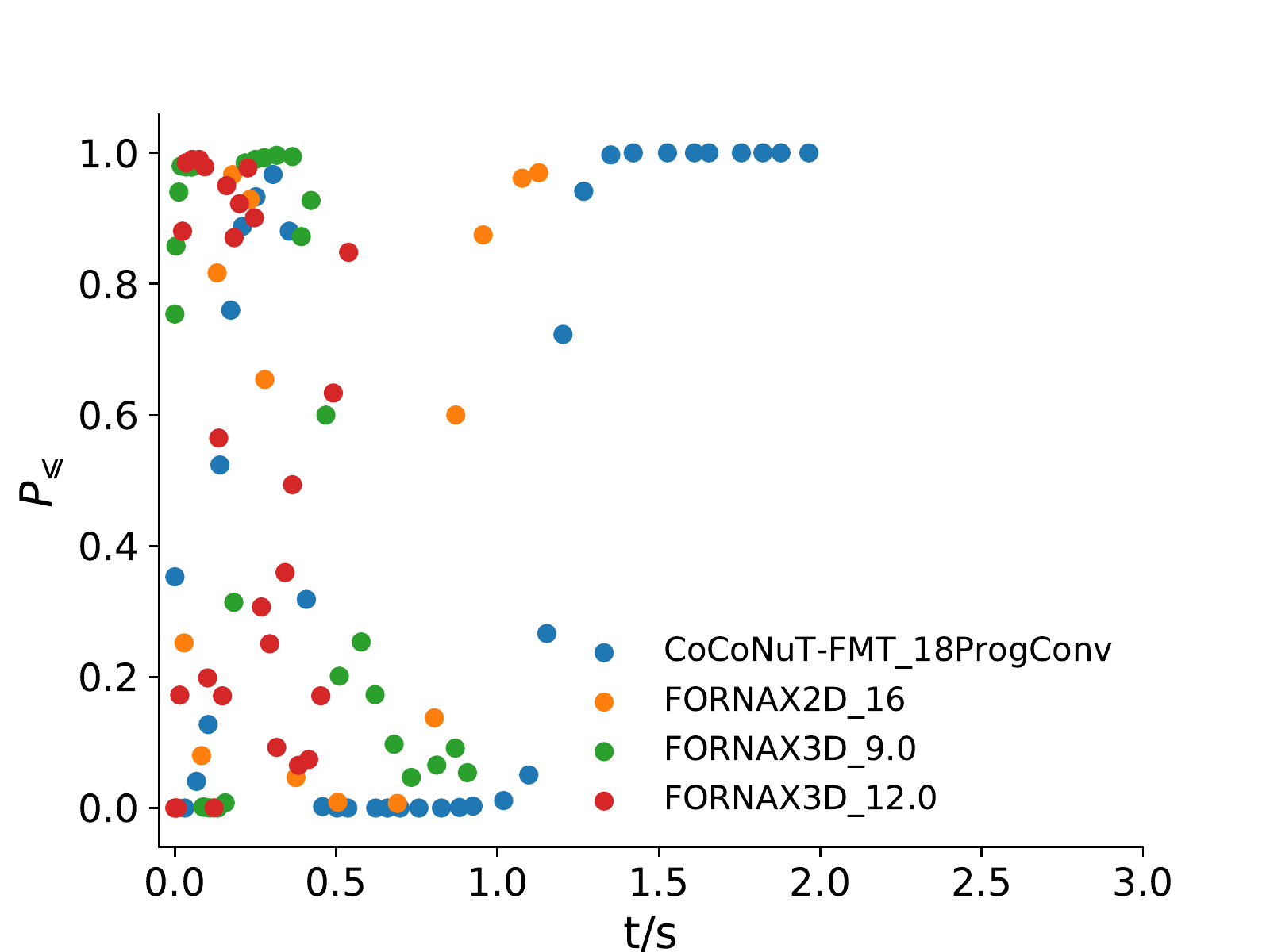}
\caption{P-value vs time for hypothesis I. For a good
  fit, the p-value should
  be randomly distributed about from 0 to 1.  The top panel shows good
	fits, and the bottom panel shows models with poor fits.
	Figures~\ref{Cumulative}~\&~\ref{KSvsR0} show the corresponding cumulative distributions
	and KS tests.}
\label{PvalueVSTime}
\end{figure}

Figure~\ref{PvalueVSTime} shows the
goodness-of-fit probability , $P_{\leqslant}$, as a function of time for hypothesis I.
Each set of dots represents the fit of one simulation.  The top panel
shows the fits which are consistent with a uniform distribution for
$P_{\leqslant}$ from 0 to 1.  The fact that there they do represent a
normal distribution is a good indicator that the model does fit the
simulation results.  The bottom panel shows a few examples
that are poor fits.  For each of these poor fits, the average value of
$P_{\leqslant}$ is around 0.5.  However, the distribution is
not uniformly distributed from 0 to 1;
the values tend to bunch up around 0 and 1, indicating that the model
is often far from the simulation data. 

\begin{figure}
\includegraphics[scale=0.5]{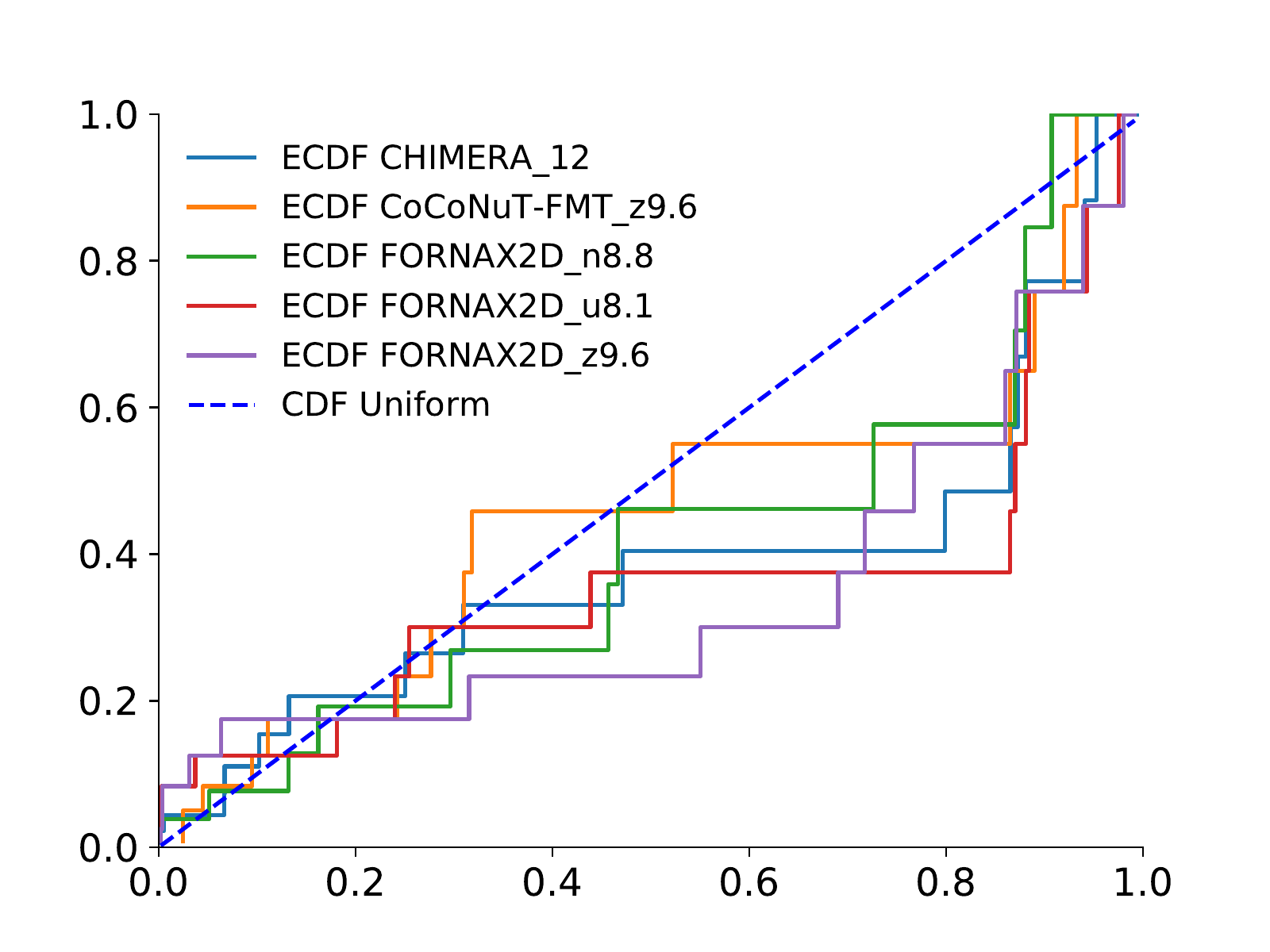}
\includegraphics[scale=0.5]{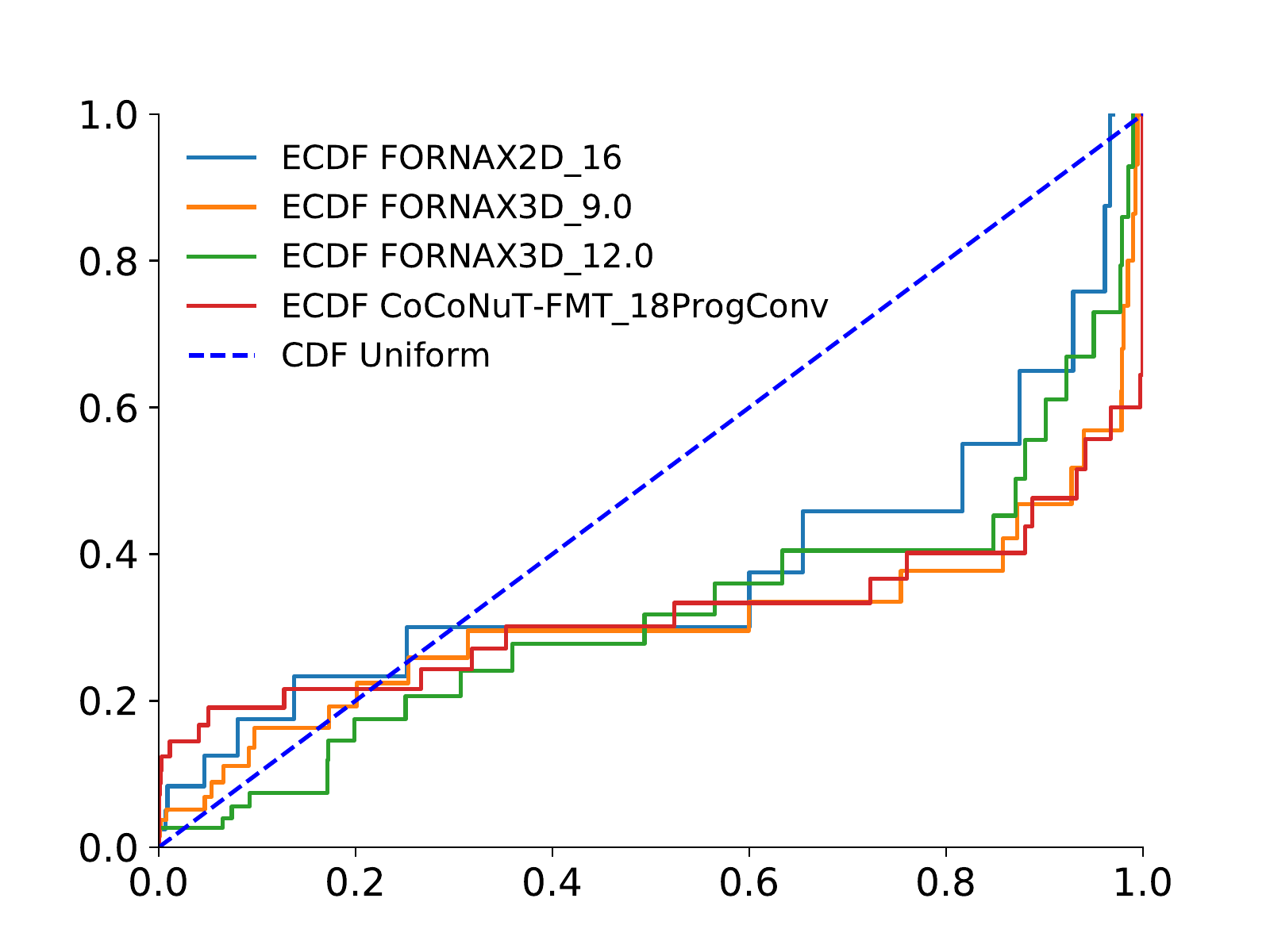}
\caption{Cumulative distribution for good (upper panel) and poor
  (lower panel) fits for hypothesis I. The solid lines   represent the P-value cumulative distributions for each simulation fit  while
  the dashed line represents a uniform distribution.  A
	good fit should have P-values that are uniformly distributed
	between 0 and 1.  The fits in the top panel are consistent with a
	uniform distribution, and
	the fits in the bottom panel are inconsistent with a uniform
	distribution.  As expected, the model is most consistent with
	simulations that explode spherically (top panel).}
\label{Cumulative}
\end{figure}

Figure~\ref{Cumulative} shows the cumulative distribution of the
goodness-of-fit probabilities for hypothesis I.  Again, each cumulative distribution
represents a fit to one simulation, the top panel shows fits which
have a uniform distribution for $P_{\leqslant}$ representing a good
fit, and the bottom panel shows the simulations which exhibit a bad
fit.  In general, core-collapse simulations of more massive stars
exhibit poor fits.  This is consistent with the expectation that the
explosion energy model of this manuscript is spherical in nature, the
low mass explosions are mostly spherical, and the higher mass
explosions are multi-dimensional. 

\begin{figure}
\includegraphics[scale=0.5]{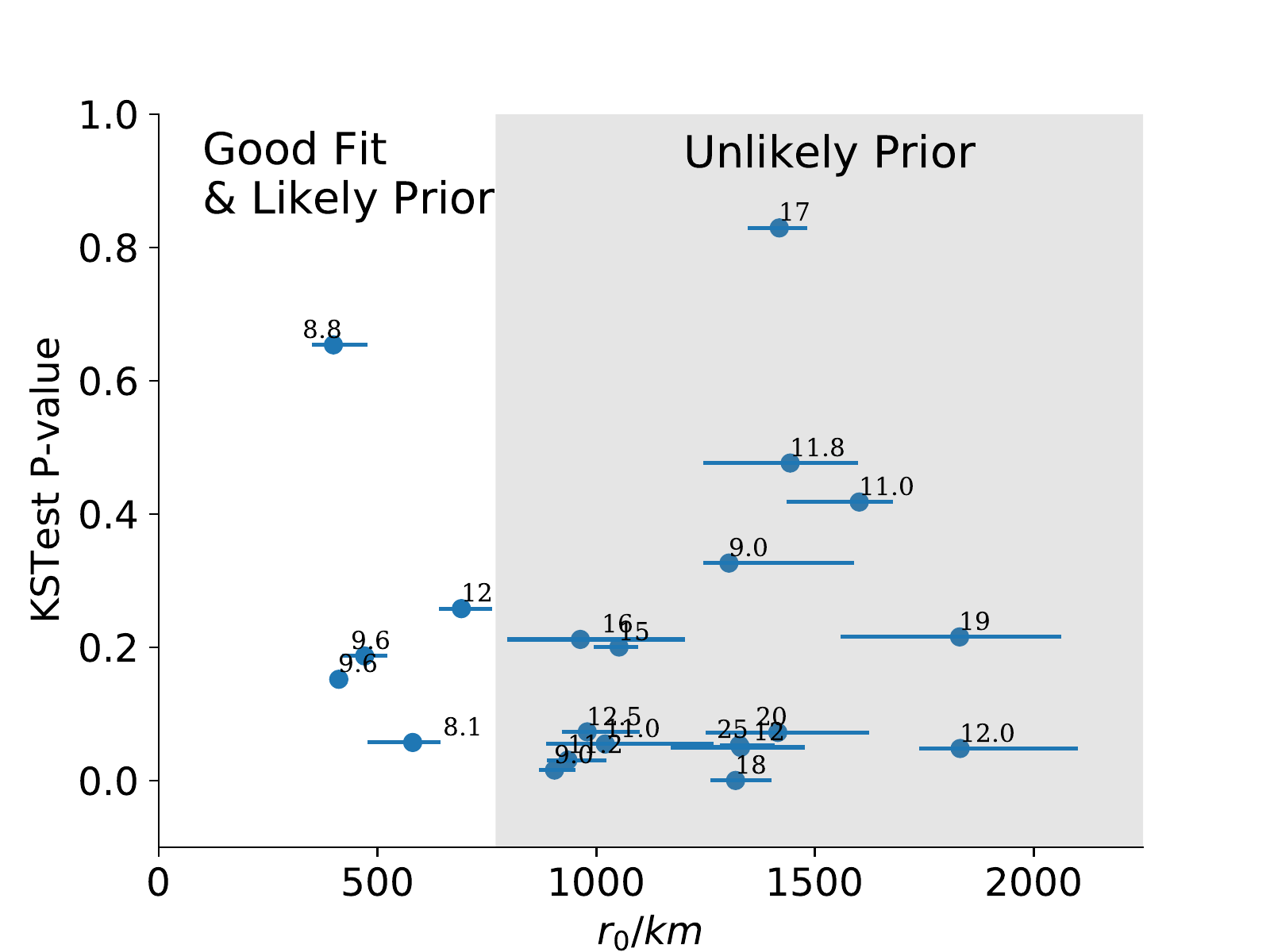}
\caption{One-sided Kolmogorov-Smirnov test result- p-value
  vs the initial shock radius, $r_0$ for hypothesis I.  Fits with high KS
	P-values are good fits by definition, but those with high $r_0$ likely do not
	represent the conditions of CCSN simulations.  Therefore, we only
	consider those with large KS P-values and $r_0$ below 800 km both
  a good fit and a reasonable representation of the CCSN simulations.
  For the five simulations that satisfy these conditions, they show
  spherical explosions.  As expected our explosion model provides
  reasonable fits for spherical explosions.  
  The other fits that do not fit or have large $r_0$ show
	aspherical explosions.}
\label{KSvsR0}
\end{figure}

Fig.~\ref{KSvsR0} shows the final results for the comparison of the theoretical model to simulations.  The left side of the figure shows fits that are a good fit and have fits that match prior expectations.  Our prior expectation is that the starting radius for the shock should be < 1000 km.  The simulations which have a good analytic fit and a good prior exhibit predominantly spherical explosions.  The right-hand-side of the figure shows simulations have an unlikely prior and a large fraction of them also exhibit a poor fit with the analytic solution.  The simulations in this region typically show aspherical explosions.  The consistency between the spherical explosion model and the mostly spherical simulated explosions bodes well for the spherical explosion model.  The fact that the spherical explosion model cannot fit the simulations with mostly aspherical explosions suggests that multi-dimensional effects are important in the explosion dynamics.

\section{Discussion and Conclusion} \label{sec4}

We present a spherically symmetric theoretical model for CCSN explosion energies. In
this model, we make the following assumptions: 1) explosion is driven by
the delayed-neutrino mechanism and $\alpha$ recombination may
  play an important role, 2) the neutrino
  luminosity is constant, 3) we follow the dynamic evolution of a spherically
  symmetric, one-zone gain region,  4) mass accretion is negligible,
  5) the gain region starts in hydrostatic equilibrium. To include
	$\alpha$ recombination energy, we developed two
	hypotheses.  I) The explosion energy is set by both
  neutrino power and $\alpha$ recombination; II) $\alpha$ recombination offsets $\alpha$
  dissociation during collapse and the explosion energy is powered by
  neutrinos alone.  Given these
  assumptions, we propose important scales for dimensional analysis,
  find solutions to the dimensionless evolution, and derive analytic
  scalings. We find  that both hypotheses 
  are consistent with the explosions of low mass progenitors which
  also exhibit mostly spherical explosions; both hypotheses are
	inconsistent with the explosion energies of the more massive
	progenitors, which also tend to explode aspherically.  This inconsistency
	suggests that the explosion dynamics of high mass
	progenitors is in part determined by multi-dimensional dynamics.
For the mostly spherical explosions, we find that the evolution of the dimensionless explosion energy, $\tilde{E}$ depends upon two dimensionless parameters, $\eta$, and $\beta$ in the case of hypothesis I and one dimensionless parameter $\eta$ in the case of hypothesis II.  Finally, we derive an analytic expression
  for the asymptotic explosion energy: $E_{\infty} \approx M_g
  \varv_0^2 \left[1.5\eta^{2/3}+\beta f(\rho_0) \right]$, where $\beta=0$ in the case of hypothesis II (neutrino heating only). 

A survey of the literature reveals three other semi-analytic or
  analytic investigations of explosion energy; however, those
  investigations had different goals and results.  In particular, 
these previous investigations did not  focus on deriving the analytic scalings nor did those
investigators directly compare their predictions with
multi-dimensional simulations.

The theoretical model in \citet{Janka2001} considered a boundary value problem
and assumed three
regions
(hydrostatic cooling, hydrostatic heating, and
free-fall layers). While this model presented both an explosion
  condition and explosion energy evolution, there were many parameters
  in the model.  As a result, it is not clear which parameters
  determine the explosion energy evolution.  Also, this study did not
  compare the theoretical explosion evolution with multi-dimensional
  simulations.

\citet{Muller2016} presented a model that considers explosion as a
two-phase process. The pre-explosion phase is described by the three-layer
model of \citet{Janka2001}.  The model for the explosion phase is
  similar to the one presented in this manuscript; it is a
  spherically-symmetric one-zone model.  However, the primary goal of
  the \citet{Muller2016} explosion model is to infer explosion
energies given a progenitor structure.  The model does not give
evolution of the explosion energy, nor does it represent the final
explosion energy in terms of $L_{\nu}$, $M_{\text{NS}}$, etc.

Moreover, \citet{Muller2016} scale the
  energy by the recombination energy of neutrons and protons into Fe
  rather than the gravitational energy of gain region.  During
  collapse, the matter loses internal energy when the Fe disassociates to free neutrons and
protons, and during explosion, matter regains that
  energy when the free baryons  recombine into bound
nuclei.  Given these facts, one might expect the dissociation and
recombination of Fe have no net effect on the explosion energy.
 \citet{Muller2016} suggest that the condition for explosion
is met when neutrinos heat the gain region enough to offset the
initial Fe dissociation. In this scenario, the explosion energy is
set by Fe recombination and the amount of mass ejected. In this paper, we investigate the role of recombination in the explosion energy evolution. We discuss two hypotheses in which explosion energy is powered by I) neutrino heating+$\alpha$ recombination and II) neutrino heating only. In difference of \citet{Muller2016} paper, in both our hypotheses neutrino heating directly contributes in the explosion energy. 
Both of the hypotheses fit low mass spherically symmetric explosions. We do not distinguish between the two.
However, \cite{Muller2016} note a correlation between explosion energy
and the Fe recombination for 2D simulations.  This prior might suggest that hypothesis I is the theory explaining energy evolution of the spherically symmetric explosions.   

We recommend further numerical investigations into whether the explosion energy depends upon the recombination.  Further detailed
analyses of numerical simulations will undoubtedly help to resolve
this mystery.
Hypothesis I predicts $E_\infty/M_g v_0^2
  \left[1.5\eta^{2/3}+\beta f(\rho_0)\right]\sim 1$, a prediction
	that is easily falsifiable by multid-dimensional simulations of
	low mass progenitors.

\citet{Papish2018} presented an order of magnitude analysis for
  the explosion energy. They found that the
delayed-neutrino mechanism can not 
yield explosions with energies more
than about $0.5 \times 10^{51} erg$. They then
conclude that the neutrino mechanism is incapable of generating
explosion energies that are inferred from observations.  However, our
analysis shows that an explosion model based upon the delayed neutrino mechanism
is consistent with the least energetic, spherically symmetric
explosions in multi-dimensional simulations.  Including aspherical
dynamics in the model could increase the explosion energy.  Therefore,
we find it premature to rule out the delayed neutrino mechanism.      

\citet{murphy2019} found that recent multi-dimensional
  simulations are under-energetic compared to the explosion energies
  inferred from observations by at least a factor of 2.  The
  resolution in this discrepancy may result from either improved
  simulations or inferences of explosion energies from observations.
  With regard to the simulations, is it even possible for the
  explosions energies to increase by more than a factor of 2?  One may use the analytic explosion
energy equation (eq.~(\ref{E_tot})) to investigate the sensitivity
of the simulated explosion energies.  However, given that this model
only explains the spherical explosions, a complete sensitivity study
must await an explosion energy theory that includes multi-dimensional dynamics.

\section*{Acknowledgements}
We thank Peter H\"{o}flich, Raphael Hix, and Frank Timmes for discussions
concerning Fe and $\alpha$ recombination.

\textit{Data Availability}: The data underlying this article are available in the article and in its online supplementary material.
\medskip
\bibliography{References}

\bsp
\appendix

\section{Evolution Equations for the One-Zone Model}
\label{appendix}

The goal of this manuscript is to derive the dimensionless variables
and analytic scalings of CCSN explosions.  Specifically, we consider
neutrino and recombination powered explosions.  Generally, the governing hydrodynamic equations, eqs.~(\ref{continuity})-(\ref{InternalEnergy}),
  are nonlinear and permit a wide range of solutions.  Sometimes,
  symmetries and constraints enable exact analytic solutions;
  other times, these constraints are not readily apparent and the
  solution seems to require numerical techniques.  Often,
  even when there are no exact analytic solutions, one may employ
  approximations, enabling analytic solution that reasonably matches the
  behavior of the numerical solutions.  In
  these situations, the challenge is finding the balance between
  the assumptions that reduce complexity while maintaining fidelity.  If the assumptions are too restrictive
  or simplifying, then the final analytic solution bears little
  resemblance with the numerical solution.  On the other hand, in an
  attempt to maintain high fidelity, one may retain too much
  complexity, and this complexity can obfuscate the essential
  physics.  As an example, a common mistake is to keep all terms in an
ODE.  Often, one may either ignore a term because other terms dominate
or combine it with another because both terms scale similarly.

In the following derivation, we start with the exact
  continuous hydrodynamic equations.  We then integrate these
  equations over the gain region, and present an exact set of one-zone evolution
  equations.  Before we make any assumptions, we define dimensionless
  variables that are likely important for the problem.  Regardless of
  whether one is able to find analytic solutions, the Buckingham $\pi$
  theorem states that all theories can only depend upon dimensionless
  variables.  The trick is to identify the most relevant and
  insightful dimensionless
  parameters.  After,
  identifying these dimensionless parameters, we then make some approximations
  that enable simpler solutions.

The first assumption is that the EoS is a gamma-law EoS, $P =
  (\gamma -1) \rho \epsilon$; using this EoS, we recast the internal
  energy equation, eq.~(\ref{InternalEnergy}), as an equation for the pressure:
\begin{equation}
\frac{dp}{dt}=-\gamma p \boldsymbol{\nabla} \cdot \boldsymbol{\varv} +(\gamma-1)\rho q_\nu+(\gamma-1) \rho q_\alpha
\label{appendp}
\end{equation}
  
To facilitate analytic solutions, our next step is to consider a
  one-zone model for the gain region. In general, the change of
  mass in the gain region is
\begin{equation}
\dot{M}_g=\frac{\partial}{\partial t} \int_{r_{\rm b,in}}^{r_{\rm
	b,out}}{\rho(r,t) dV} \, ,
\end{equation}  
where $r_{\rm b, in(out)}$ are the inner (outer) boundaries of the
gain region. In this model, the inner
boundary is fixed, while the outer boundary changes in
time \citep{Vartanyan2019}. Using Leibniz's integral rule, the change in mass is:
\begin{equation}
\dot{M}_g= 4\pi r_{\rm b,out}^2 \rho(r_{\rm b,out},t) \varv_{\rm b,out}+ \int_{r_{\rm b,in}}^{r_{\rm b,out}}{\frac{\partial \rho}{\partial t} dV} 
\label{mdotmid}
\end{equation}
The second term of eq.~(\ref{mdotmid}) can be rewritten using the
continuity equation (eq.~(\ref{continuity})) and Gauss' theorem:
\begin{equation}
\dot{M}_g= 4\pi r_{\rm b,out}^2 \rho(r_{\rm b,out},t) \varv_{\rm b,out} - \int_{r_{\rm b,in}}^{r_{\rm b,out}}{\rho \boldsymbol{\varv} \cdot \boldsymbol{dS}} 
\label{exactmdot}
\end{equation}
In addition to assuming
  that the inner boundary is stationary, we also assume that the fluid
velocity at the inner boundary is zero \citep{Liebendorfer2001b}.
Given these assumptions, the mass change is
\begin{equation}
\dot{M}_g=4\pi r_{\rm b,out}^2 \rho(r_{\rm b,out},t) (\varv_{\rm b,out}- \varv_{\rm fluid}) \,
\label{finalmdot}
\end{equation}
where $\varv_{\rm fluid}$ is the velocity of the fluid at the outer
boundary. This difference in the boundary velocity (the shock)
  and the fluid velocity represents the mass flux through the shock
  into the gain region.
  In the strong shock limit, the shock compression is a single factor
  and the shock (boundary) velocity is a constant multiple of the
  post-shock fluid velocity. For the range of
	possible $\gamma$, shock strength, and recombination, the shock
	compression can range from a factor of 5 to 10.  Regardless, the
	post shock fluid velocity will be a multiple of the shock
	velocity.  Therefore, all velocities will scale similarly, and we choose one representative velocity to represent the evolution
	of all velocities $\varv_c$.
  
  Furthermore, we note that the accretion timescale is much
	longer than the dynamical timescale. For 
  typical values of the initial shock radius, $r_0=150$
  km \citep{Liebendorfer2001b}, the dynamical timescale is $t_{\rm D}\sim 0.003$ s, and the
  accretion timescale is $t_{\rm acc}=M_g/\dot{M}\sim 0.2$
  s. Since the explosion timescale is a few dynamical
	timescales, the explosion timescale is also shorter than the
	accretion timescale.  Given these results, we assume that
	accretion does not change the mass in the gain region during
	explosion.  However, the explosion timescale is only a
	factor of few shorter than the accretion timescale.  Therefore, we
  cautiously proceed with this assumption.
   We choose to proceed
	with this assumption because it greatly reduces the complexity of
  the equations. If the model does not
	match the simulations,
  then this would be one of the first assumptions to
  investigate. The fact that the model is consistent with
	spherically symmetric simulations (section \ref{sec3}) suggests that
	this assumption is not grossly incorrect. Therefore,
  we assume that one characteristic velocity
represents the expanding shell and shock, and the gain region mass does
not change much; $\dot{M}_g \approx 0$.

The one-zone, volume-averaged equations for momentum and pressure are:
\begin{equation}
\frac{1}{V_g} \int_{V_g}{\rho \frac{d \varv}{dt}dV}=-\frac{1}{V_g}\int_{V_g}{\nabla_r p dV}-\frac{1}{V_g}\int_{V_g}{\rho \nabla_r \Phi dV} \,
\label{appintdv}
\end{equation}

\begin{align}\label{appintdp} 
\frac{1}{V_g}\int_{V_g}{\frac{dp}{dt} dV}=& -\frac{\gamma}{V_g}\int_{V_g}{p \boldsymbol{\nabla}\cdot \boldsymbol{\varv} dV}+\frac{\gamma-1}{V_g}\int_{V_g}{\rho \frac{L_\nu \kappa}{4 \pi r^2}dV} \\ \nonumber
&+\frac{\gamma-1}{V_g}\int_{V_g}{\rho q_\alpha dV} \, 
\end{align}
where $V_g$ is the volume of the gain region. While these equations are
exact, the solutions are not necessarily analytic.  To facilitate
analytic solutions, we next impose some approximations and constraints.

Many of the terms in the one-zone momentum and pressures
  equations have at least two variables within an integral.  To
  simplify these terms, we use the second mean value theorem for definite integrals:
\begin{equation}
\int_{x_1}^{x_2} f(x)g(x)dx=f(x')\int_{x_1}^{x_2} g(x)dx \,
\label{theorem}
\end{equation}
where $f(x)$ and $g(x)$ are some arbitrary, intagrable, continuous
functions, and $x'$ lies within $x_1$ and $x_2$,      
 $x_1\leq x' \leq x_2$.  This approximation is
  effective as long as $f(x^{\prime})$
  is similar to $f(x_1)$ and $f(x_2)$.  For example, this
  approximation has the correct order if $f(x^{\prime})$, $f(x_1)$,
  and $f(x_2)$ are all of
  the same order, or the approximation has the correct scaling if they
all have the same scaling.  As an example, under the second mean value
thereom, the left term of eq.~(\ref{appintdv})
becomes $\frac{1}{V_g} \frac{d\varv}{dt}(r^{\prime}) \int \rho
\, dV$.  The integral of the density is the mass in the gain region,
$M_g$.  In general, any of the radial velocities within the gain region
are of the same order during explosion both in magnitude, direction,
and change.  Furthermore, the postshock velocity is of order the shock
velocity because they are connected by a constant factor.  Hence,
$\frac{d\varv}{dt}(r^{\prime})$ roughly represents the time evolution
of any velocity
associated with the explosion, including the shock velocity.  We
choose a characteristic velocity to represent all of these,
$\varv_c$.  Finally, we have an approximation for the acceleration of
the gain region, $\frac{M_g}{V_g} \frac{d \varv_c}{dt}$.

Again, we use the second mean value thereom,
  eq.~(\ref{theorem}), to approximate the
  pressure gradient term on the right side of eq.~(\ref{appintdv}).  This
  pressure gradient term is approximately $-\frac{1}{V_g}4 \pi r_c^2 \int{\frac{\partial p}{\partial
	r}dr}$.  Evaluting this integral, the pressure gradient
  term is approximately $4\pi r_c^2 (p_{\rm in}-p_{\rm out})$.  $p_{\rm in}$
  represents the pressure at the inner radius of the gain region, and
  $p_{\rm out}$ represents the pressure at the outer radius of the
  gain region.  The latter is set by the ram pressure.  Since
	$p_{\rm in}$ is much larger than $p_{\rm out}$ (see below), the
	net pressure gradient force on the zone points outward.

Next, we show
  that the pressure gradient term can be approximated as
  $\frac{1}{V_g}4\pi r_c^2 p_c$, where $p_c$ is a characteristic
  pressure scale.  First, we note that while the ram pressure acts as
  an inward force on the gain region, it is not the primary reason
  that the shock does not expand.  It is true that the ram pressure
  helps to determine the location of the stalled shock before
  explosion. However, the energetics and dynamics of the gain region
  are mostly set by hydrostatic balance between pressure and
	gravity, neutrino heating, and advection. Before explosion, the post shock region
  is in near hydrostatic balance, and it remains that way because
  while neutrinos heat the gain region, advection quickly advects that
  material downward to the cooling region.  In essence, the shock is stalled
  because advection of internal energy downward balances the heating
  by neutrinos in the gain region.  The situation explodes when the
  downward advection can no longer balance the neutrino heating,
	and the global pressure gradient overwhelms gravity.
  For a more thorough derivation of this fact, see
 \cite{murphy2017} and references therein.
  Once this delicate balance tips toward explosion, the energy
  scale and forces within the gain region quickly overwhelm the ram
  pressure.

To quantitatively illustrate this, consider the
  following calculations.  The ram pressure of the infalling material
can be estimated as $p_{\rm ram}\sim \rho_{\rm out} \varv_{\rm ff}^2$,
where $\varv_{\rm ff} \sim\sqrt{GM_{\rm NS}/r}$ is the free-fall
velocity and $\rho_{\rm out}$ is the density of the material outside
of the shock. $\rho_{\rm out }$ can be determined from the mass conservation: $\rho_{\rm out}=\frac{\dot{M}}{4\pi r^2 \varv_{\rm ff}}$. Thus, the ram pressure is 
\begin{equation}
p_{\rm ram}\sim \sqrt{\frac{G M_{NS}}{r}} \frac{\dot{M}}{4\pi r^2} 
\end{equation}  
A dimensionless scale for ram
pressure is $\tilde{p}_{\rm ram}\sim p_{\rm
  ram}/p_0$, where $p_0$ is an initial condition for the
  characteristic pressure, $p_0=\frac{G M_{\rm NS} M_g}{4\pi r_0^4}$. For typical values of physical parameters: $M_g=0.05M_\odot$, $M_{\rm NS}=1.4 M_\odot$, $r_0=150$km, $\dot{M}=0.25 M_\odot s^{-1}$, we obtain:
\begin{equation}
\tilde{p}_{\rm ram} \sim \frac{2\times 10^{-2}}{\tilde{r}^{5/2}} \lesssim 2 \times 10^{-2}
\end{equation}
In comparison, the dimensionless characteristic pressure
at $\tilde{t}=3.5,$  is $\sim
  0.2$ (Figure~\ref{Pteta}). At this time and radius, the dimensionless ram pressure is $\sim
0.5 \times 10^{-2}$. Therefore, $\tilde{p}_{\rm ram}/\tilde{p} \sim
2.5 \%$.  Hence, $P_{\rm out}$ is a small correction to the
  hydrostatic balance established in the gain region. 

Furthermore, we illustrate that neutrinos quickly overwhelm the ram pressure
during explosion.
To do so, we estimate the change in pressure due to neutrino heating
  during the explosion. The dynamics of dimensionless pressure is given by eq.~(\ref{dpdttilde}).  First, we integrate the change of dimensionless pressure due to neutrino heating (second term of eq.~(\ref{dpdttilde})) with dimensionless time and use the asymptotic
relation for the shock radius (eq.~(\ref{Rassymptot})). The pressure
change due to neutrino heating $\Delta \tilde{p}_\nu$ reaches the ram
pressure $\tilde{p}_{\rm ram}$ in $\tilde{t}\sim 0.2$. Neutrinos
continue to increase the pressure up until $\tilde{t}\sim 4$
 making the change in pressure due to
  neutrinos a factor of $10$ larger than the initial ram pressure.

Taken together, the negligible contribution of the ram pressure
  to the overall pressure scale and the fact that added pressure due
  to neutrino heating quickly overwhelms the ram pressure, we ignore the effect of ram
pressure in the gradient term.

To approximate the gravity term, the last term in
eq.~(\ref{appintdv}), we again
use the second mean value theorem for definite
integrals. Using this approximation,
  the gravity term is approximately  $-\frac{G M_{NS}}{r_c^2
  V_g}\int{\rho dV}$, or $-\frac{G
  M_{NS}}{r_c^2}\frac{M_g}{V_g}$ for the gravitational force.  This completes the approximate momentum equation:
\begin{equation}
\frac{d \varv_c}{dt}=-\frac{G M_{NS}}{r_c^2}+\frac{4 \pi r_c^2
  p_c}{M_g} \, .
\label{appendvc}
\end{equation}

Next, we simplify the internal energy or pressure equation,
eq.~({\ref{appintdp}}).  For the left-hand side, we note that
  the derivative inside the integral is a Lagrangian derivative, and
  as long as the boundaries are fixed with respect to the Lagrangian
  mass, then one may pull out the Lagrangian time derivative.  One is
  then left with an average or characteristic pressure for the gain
  region.  Hence, the left-hand side is approximately, $dp_c/dt$.
 For the first
term on the right-hand side of eq.~({\ref{appintdp}}), we first
use the mean-value theorem (\ref{theorem}) to bring the
  pressure out of the integral:  $-\frac{\gamma}{V_g}
p_c \int{\boldsymbol{\nabla} \cdot
  \boldsymbol{\varv}dV}$. Then Gauss's
theorem and a few approximations give $-4\pi r_c^2 p_c \varv_c
\frac{\gamma}{V_g}$. Technically, application of Gauss' Theorem results in
terms involving both
  the bottom and top of the gain region.  However, the bottom is
  stationary in this model, and we note that all velocities are of
  order $\varv_c$ (see discussion above).

Next, we show that the volume scales as $V_g \sim r_c^3$, and
  more specifically, we approximate it as $V_g\sim \frac{4}{3}\pi r_c^3$.
  To be precise, the volume of the gain region is $V_g=\frac{4}{3}\pi (r_{\rm
   out}^3-r_{\rm in}^3)$. If the inner radius remains fixed, then
	the volume quickly becomes dominated by the $r_{\rm out}^3$ term.
    Hence, one may ignore $r_{\rm in}^3$ and assume that
 the outer radius scales as the characteristic radius of the gain
 region. Even if one considers a thin shell approximation, the
   thickness of this shell will likely scale as $\Delta r \sim r_c$.
 Therefore, regardless of the approximation, the volume will
   scale as $V_g \sim r_c^3$.

 Given these approximations, we propose approximations for the three
 right-hand side terms in the pressure equation,
 eq.~(\ref{appintdp}).  The
  adiabatic cooling term, the first term on the right-hand side, is
  approximately $-3\gamma\frac{p_c\varv_c}{r_c}$.
 To approximate the neutrino heating term, the second term of eq.~({\ref{appintdp}}), we use similar approximations
  as in the gravitational force term, yielding 
$(\gamma-1)\frac{M_g}{V_g}\frac{L_\nu \kappa}{4\pi r_c^2}$. Lastly,
for the recombination energy term of eq.~({\ref{appintdp}}),
 the mean value theorem gives $(\gamma-1)q_{\alpha c}\frac{M_g}{V_g}$.
Thus, the approximate form of dynamic of pressure is 
\begin{equation}
\frac{dp_c}{dt}=-3\gamma \frac{p_c \varv_c}{r_c}+\frac{(\gamma-1)L_\nu
  \kappa}{\frac{16}{3} \pi^2 r_c^5}M_g+\frac{3M_g}{4 \pi r_c^3}
(\gamma-1)q_{\alpha c} \, .
\label{appendpc} 
\end{equation}
For convenience, we drop the subscript ``c" in
eqs.~(\ref{appendvc})-(\ref{appendpc}), which
leads  to eqs.~(\ref{dvdt})-(\ref{dpdt}).

\label{lastpage}
\end{document}